\documentclass[preprint]{aastex}
\usepackage{epsfig}
\usepackage{amssymb}
\usepackage{makeidx}
\usepackage{ifthen}
\usepackage{verbatim}
\begin{document}

\newcommand{\be}{\begin{equation}}
\newcommand{\ee}{\end{equation}}
\def\rf{rest frame}
\def\lcf{light curve fitter}
\def\rs{redshift}

\title{Reducing Zero-point Systematics in Dark Energy Supernova Experiments}
\author{L. Faccioli\altaffilmark{1}, A. G. Kim\altaffilmark{2}, R. Miquel\altaffilmark{3}, G. Bernstein\altaffilmark{4}, A. Bonissent\altaffilmark{5}  M. Brown\altaffilmark{6}, W. Carithers\altaffilmark{2}, J. Christiansen\altaffilmark{7}, N. Connolly\altaffilmark{8}, S. Deustua\altaffilmark{9}, D. Gerdes\altaffilmark{10}, L. Gladney\altaffilmark{4}, G. Kushner\altaffilmark{2}, E. V. Linder\altaffilmark{1,11,12}, S. McKee\altaffilmark{10}, N. Mostek\altaffilmark{1}, H. Shukla\altaffilmark{2} A. Stebbins\altaffilmark{13}, C. Stoughton\altaffilmark{13}, D. Tucker\altaffilmark{13}}

\altaffiltext{1}{Space Sciences Laboratory, University of California at Berkeley, Berkeley, CA 94720, USA}
\altaffiltext{2}{Lawrence Berkeley National Laboratory, Berkeley, CA 94720, USA}
\altaffiltext{3}{Instituci\'o Catalana de Recerca i Estudis Avan\c{c}ats, Institut de F\'{\i}sica d'Altes Energies, E-08193 Bellaterra (Barcelona), Spain}
\altaffiltext{4}{Department of Physics \& Astronomy, University of Pennsylvania, Philadelphia, PA 19104, USA}
\altaffiltext{5}{CPPM-IN2P3-CNRS, Campus de Luminy, 13288 Marseille CEDEX 9, France} 
\altaffiltext{6}{Lincoln Laboratory, Massachusetts Institute of Technology, Lexington, MA 02420, USA}
\altaffiltext{7}{Physics Department, California Polytechnic State University, San Luis Obispo, CA 93407, USA}
\altaffiltext{8}{Physics Department, Hamilton College, Clinton, NY 13323, USA}
\altaffiltext{9}{Space Telescope Science Institute, Baltimore, MD 21218, USA}
\altaffiltext{10}{Physics Department, University of Michigan, Ann Arbor, MI 48109, USA} 
\altaffiltext{11}{Berkeley Center for Cosmological Physics, Berkeley, CA 94720, USA}
\altaffiltext{12}{Institute for the Early Universe, Ewha University, Seoul, Korea}
\altaffiltext{13}{Fermi National Accelerator Laboratory, Batavia, IL 60510, USA}
\begin{abstract}
We study the effect of filter zero-point uncertainties on future supernova dark energy missions.
Fitting for calibration parameters using simultaneous analysis of all Type Ia supernova standard candles
achieves a significant improvement over more traditional fit methods.
This conclusion is robust under diverse experimental configurations
(number of observed supernovae, maximum survey redshift, inclusion of additional systematics).
This approach to supernova fitting considerably eases otherwise stringent mission calibration requirements.
As an example we simulate a space-based mission based on the
proposed JDEM satellite; however the method and conclusions
are general and valid for any future supernova dark energy mission, ground or
space-based.
\end{abstract}
\section{Introduction}
\label{sec:intro}
The discovery of the acceleration of the expansion of the universe \citep{riess98,perlmutter99}
ranks as one of the most significant recent discoveries in cosmology.
This acceleration is usually ascribed to a mysterious ``dark energy'' about which almost
nothing is known although there are many competing ideas;
what is needed to distinguish between them and shed more light on the nature of the acceleration 
is more and improved data.
Observations of type Ia supernovae (SNe Ia) have allowed the discovery of the acceleration of the expansion
\citep{riess98,perlmutter99} and are currently the most established and best \citep{albrecht06}.
The method is described by many authors \citep{perlmutter97,riess98,perlmutter99,perlmutter03}
and is based on the fact that SNe Ia are, to good accuracy, standardizable candles (for a review of
SNe Ia as standardizable candles see \citet{phillips03,branch92}).
However current supernova observations are limited by systematic uncertainties; while this was not a
problem when the supernova sample was small and statistical uncertainties were the dominant ones, the growing
sample size has already reached the point when statistical and systematic uncertainties are of comparable magnitude, as in e.g.
the combined sample of $557$ supernovae studied by \citet[The Union2 compilation:][]{amanullah10}.
As more supernovae will be discovered in the future the need to better characterize and reduce systematic uncertainties will become \emph{the}
dominant concern in dark energy experiments.
This has been recognized for several years, and the SuperNova/Acceleration Probe (SNAP) 
satellite\footnote{\url{http://www.snap.lbl.gov/}}\citep[SNAP Collaboration:][]{aldering04}  was proposed as a
systematics-controlled space-based experiment that would put much tighter constraints on dark energy
than current and near future experiments by following $\approx 2000$ supernovae out to
$z_\mathrm{max}\approx 1.7$.
More recently NASA and the Department of Energy have announced the Joint Dark Energy Mission
(JDEM) \footnote{\url{http://jdem.gsfc.nasa.gov/}}$^,$\footnote{\url{http://jdem.lbl.gov/}}
as a future space-based mission to study the nature of dark energy by employing a combination of techniques including supernovae.
Therefore it is important to characterize the sources of systematics of future supernova experiments;
studies of this kind have already appeared \citep{kim04, kim06, nordin08, linder09} and this paper aims at building upon and expanding them.
Studies using real data-sets have also appeared: for example \citet{kilbinger09} use 
the SNLS supernovae in \citep{astier06} to evaluate the effect of zero-point
uncertainties on the final cosmology.
\par
The two most important sources of systematic uncertainty in dark energy experiments that use supernovae are
the dimming by dust in the host galaxy and uncertainties in the flux calibration, specifically the filter zero-points,
as seen in recent cosmological analyses such as \citet{astier06}, \citet{woodvasey07}, \citet{kowalski08}, \citet{hicken09}, \citet{amanullah10}.
The problem of host-galaxy dimming is also being aggressively pursued, by e.g. targeting supernovae in rich clusters of galaxies
\citep{dawson09}; we will include it statistically in our analysis but will not go into its systematics.
Properly taking into account zero-point uncertainties is nontrivial because their causes are numerous, ranging from any
inaccuracy in the response function of telescope, filter, or detector (from now on collectively indicated as
``channel''), or the atmosphere for ground-based experiments, to uncertainties in the calibration procedure.
While accurately characterizing all these is obviously an experiment-dependent problem, our aim is to
provide a more general way to deal with them.
\par
Starting with a simple model of zero-point uncertainty, we perform a complete
end-to-end simulation of a supernova dark energy mission, propagating zero-point uncertainties through the
simulation chain, and we evaluate its effects on the final cosmology fit. 
We do not aim at a detailed physical modelling of particular causes of uncertainty
such as imperfect knowledge of the standard stars used to calibrate the zero-points or of the filter response functions,
but rather at characterizing their overall effect, whatever their underlying reasons, with a set of zero-points
representing the contribution of these important sources of systematics to the final error budget.
This will serve as a guide to designers of how much specific components (telescope, filters, detectors,
calibration procedure and so on) could be imperfectly known and still achieve the mission objectives.
\par
Our starting point is the result by \citet[][hereafter KM]{kim06}.
KM introduce a new model of filter zero-point uncertainty and show that, due to the standardizable candle nature of
SNe Ia, it is possible to treat these uncertainties as parameters that can be included with other parameters
in a cosmology analysis.
More precisely, KM model the observed peak magnitude $m$ of a supernova as 
$m = \mu + M + \mathit{Other} + \mathcal{Z}$ where
$\mu$ is the distance modulus, $M$ is the absolute magnitude after standardization
(and therefore the same for all supernovae), ``$\mathit{Other}$'' indicates all residual effects that
influence $m$, such as host galaxy extinction, and $\mathcal{Z}$ is a new 
fit parameter for zero-point offsets to be fit with all the other parameters in the model.
It is important to note that modelling zero-point offsets as fit parameters would not be possible if
SNe Ia were not standardizable candles because $M$ would not be the same for every supernova. 
KM show that by fitting for all the supernovae distance moduli \emph{simultaneously}
it is possible to achieve a significant reduction in the final uncertainties in the cosmological parameters
with respect to the traditional case when supernova distances are fit one by one
and calibration uncertainties are then included in the total error budget.
In the rest of the paper we will refer to the KM fitting approach as ``simultaneous fit'' and to the traditional approach
as ``SN by SN fit''.
\par
This work expands KM in several ways:
\begin{enumerate}
\item
KM carry on a Fisher matrix analysis of their model; we perform a complete end-to-end
simulation of a supernova dark energy mission, with a realistic modelling of all its aspects.
\item
KM use a particular $z$ distribution, in which all supernovae are placed at those special redshifts
that have zero $K$-correction.
At those $z$ the improvement in mission performance is maximum; we study a more realistic $z$ distribution.
\item
We include an intrinsic color dispersion.
\item
We investigate whether our conclusions are robust with respect to several
changes in mission parameters (number of supernovae, maximum redshift, inclusion of additional
systematics); our simulation tool allows us to explore a much wider parameter space than KM.
\end{enumerate}
It is important to point out that the KM model that we adopt here is applicable to
a generic future dark energy mission based on supernovae; however, for concreteness we will
present our results by considering a specific example: the supernova survey of
the future space-based dark energy mission based on the proposed SNAP satellite.
We will also assume that a nearby sample of supernovae, whose characteristics are based on the
expected Nearby Supernova Factory sample\footnote{\url{http://snfactory.lbl.gov/}} \citep{aldering02,copin06},
is available: specifically this sample is comprised of $316$ supernovae with
$0.03\leq z\leq 0.08$.
\par
The paper is organized as follows: Section \ref{sec:snmodel} describes the KM model
and its implementation in our simulation tool, Sections \ref{sec:results} and
\ref{sec:parspace} describe our results and Section \ref{sec:conclusion} summarizes our
conclusions and discusses ways the work can be expanded.
In what follows we will use the terms ``mission'' and ``experiment'' interchangeably.
\section{Supernova model and implementation}
\label{sec:snmodel}
Our analysis begins with a set of supernova statistics (peak magnitudes and stretches) in different bands, representing both
the distant sample that our simulated mission will observe and the nearby sample that we assume already available;
these statistics are obtained in the following way.
For each supernova a redshift is chosen from a specified redshift distribution and fluxes and their uncertainties are computed
by convolving spectral templates from \cite{hsiao07} with the channel throughputs; this results in a set of simulated supernova fluxes
in different bands at different epochs.
The flux variability due to Poisson noise is simulated by drawing the fluxes from a Gaussian distribution,
which is appropriate in the limit of high expected numbers of photons.
Each band is then fit independently of the others, following \citet{perlmutter99}, to give, among other parameters,
an estimated flux at maximum and a stretch in each band; the covariance matrix is also computed and propagated later.
Up to and including the stage of light curve fitting, the two approaches, simultaneous fit and
SN by SN fit, do not differ and they are carried out in the same way by our analysis.
We then fit for the distance moduli (hereafter ``$\mu$ fit''); this is the step where the differences in the two approaches
are manifest, and we describe the models we used for each approach in more detail in the next two subsections.
We then describe the data error model used both by the $\mu$
fit and the cosmology fit, the cosmology fit itself, which again is performed in the same
way for simultaneous fit and SN by SN fit, and finally the mission parameters we use.
In the rest of the paper $N_\mathrm{SN}$ will denote the number of supernovae observed by the mission,
excluding the nearby sample and $N = N_\mathrm{SN}+N_\mathrm{Nearby}$ will denote the total number
of observed supernovae.
\subsection{Simulating the zero-point uncertainty in the SN by SN fit: a Monte Carlo approach}
\label{sec:mc}
In this section we present how the SN by SN analysis is performed.
After converting fitted peak fluxes to magnitudes we model these and the stretches as:
\begin{eqnarray}
\label{eq:snmodel}
m^i{}_k &=& \mu^i + \alpha(S^i - 1) + M(z_i)_k + A_V{}^i a(z_i)_k + B_V{}^i b(z_i)_k\nonumber
\\
s^i{}_k &=& S^i,
\end{eqnarray}
with $i=1\cdots N$; for each supernova $i$, $k$ belongs to the subset of $\{1\cdots N_\mathrm{F}\}$
that covers restframe optical and near infra-red (NIR) wavelengths
and $N_\mathrm{F}$ denotes the number of filters used in the mission; in our simulations we assume $N_\mathrm{F}=8$.
The meaning of the symbols in Equations \ref{eq:snmodel} is as follows, distinguishing between \emph{input data}, \emph{model parameters}, and
\emph{known functions}.
\begin{enumerate}
\item
Input data from simulations:
\begin{enumerate}
\item
$m^i{}_k$ denotes the simulated peak instrumental magnitudes of supernova $i$ in observer frame band $k$ obtained
after light curve fitting.
\item
$s^i{}_k$ denotes the stretch of supernova $i$ in observer frame band $k$ after light curve fitting.
\end{enumerate}
\item
Model Parameters:
\begin{enumerate}
\item 
$\mu^i$ denotes the distance modulus of supernova $i$.
\item
$S^i$ is a weighted stretch parameter for supernova $i$ used to fit for $\mu^i$; unlike $s^i{}_k$ which depends on
the observer frame band, there is a single parameter $S^i$ for each supernova.
\item
$A_V{}^i$ and $B_V{}^i\equiv(A_V / R_V)^i$ are extinction parameters \citep[CCM:][]{cardelli89}.
\end{enumerate}
\item
Known functions:
\begin{enumerate}
\item
$M(z_i)_k$ is the absolute peak magnitude of a $S=1$ supernova at redshift $z_i$ in observer frame band $k$, given by:
\begin{equation}
\label{eq:M}
M(z_i)_k=-2.5\log\bigg(\int\mathrm{d}\lambda f((1+z_i)\lambda)T_k(\lambda)\bigg)
\end{equation}
where $f(\lambda)$ is a template spectrum from \citet{hsiao07} and $T_k(\lambda)$
is the throughput of channel $k$, with $\lambda$ the observer frame wavelength.
\item
$a(z_i)_k$ and $b(z_i)_k$ model host galaxy extinction and
are computed in a manner similar to $M(z_i)_k$:
\begin{eqnarray}
\label{ref:abccm}
a(z_i)_k&=&-2.5\log\bigg(\int\mathrm{d}\lambda a_\mathrm{CCM}((1+z_i)\lambda)f((1+z_i)\lambda)T_k(\lambda)\bigg) \nonumber
\\
b(z_i)_k&=&-2.5\log\bigg(\int\mathrm{d}\lambda b_\mathrm{CCM}((1+z_i)\lambda)f((1+z_i)\lambda)T_k(\lambda)\bigg)
\end{eqnarray}
where $a_\mathrm{CCM}(\lambda)$ and  $b_\mathrm{CCM}(\lambda)$
are known functions of wavelength, describing host galaxy extinction;
we assume a CCM extinction law.
\item
$\alpha$ is a fixed dimensionless constant; we assume $\alpha = -1.7$.
\end{enumerate}
\end{enumerate}
For each supernova $i=1\cdots N$ we fit for $\mu$, $S$, $A_V$, and $B_V$.
\par
Our chosen value for $\alpha$ was based on older values and is somewhat larger, in absolute value, than those found by recent analyses
of supernova data:
for example \citet{kowalski08} find $\alpha = -1.46 \pm 0.16$
(note that with our sign convention in Equations \ref{eq:snmodel} $\alpha < 0$) for the supernovae in the $z>0.2$ Union subsample,
which is the relevant one since in our subsequent analyses we will assume $z>0.3$.
This may result in a conservative parameter estimation in all our simulations but would not change our conclusions.
\par
We now include zero-point uncertainties, which are not described in the system of Equations \ref{eq:snmodel}.
The usual approach to incorporating zero-point uncertainties is to estimate them
and include them in the total error budget (see \citet{amanullah10} for an attempt at jointly modelling zero-point uncertainties
and other systematics taking their covariances into account).
Following KM we implement the usual approach by modelling the zero-point uncertainty in each 
bandpass $k$ via a peak magnitude shift, described by a parameter 
$\mathcal{Z}_k$, for supernova $i$ in observer frame band $k$.
\be
\label{eq:magshift}
m^i{}_k \rightarrow m^i{}_k + \mathcal{Z}_k,~\forall i
\ee
where $\mathcal{Z}_k$ is a random shift drawn from a Gaussian distribution
with $0~\mathrm{mag}$ mean; the value of its standard deviation quantifies our prior knowledge of the filter zero
point uncertainty, $\sigma_\mathcal{Z}$.
Since the \lcf~fits for a peak flux, $f_{0k}=10^{-0.4m_k(\mathrm{max})}$, the magnitude shift is actually 
converted to flux before being applied, according to the usual expression:
\be
\label{eq:magtoflux}
f_{0k} \rightarrow f_{0k} \times 10^{-0.4 \mathcal{Z}_k.},
\ee
The same magnitude shift $\mathcal{Z}_k$ affects all supernovae that are
observed through filter $k$; since this band in the restframe varies from supernova to supernova
depending on their redshifts, the same $\mathcal{Z}_k$ affects different supernovae
in different ways, introducing a \emph{correlation} between their distance moduli $\mu$.
Neglecting for the moment other sources of variability, for supernova $i$ we may write: 
$\mu^i = m^i{}_k - M(z_i)_k$, where $M(z_i)_k$ is defined in Equation \ref{eq:M}.
Then as $m^i{}_k \rightarrow m^i{}_k + \mathcal{Z}_k \Longrightarrow
\mu^i \rightarrow \mu^i + \mathcal{Z}_k$; the $\mu$s become 
correlated via the $\mathcal{Z}_k$ parameters and their covariance matrix 
becomes non-diagonal.
\par
The contribution of the filter zero-point uncertainty to the overall $\mu$
covariance matrix is estimated via a Monte Carlo approach (MC): at each MC
realization a different set of magnitude shifts $\mathcal{Z}_k$, one for
each filter, is drawn from a Gaussian distribution with $0~\mathrm{mag}$ mean and a chosen
standard deviation; we run the MC with standard deviations ranging from
$0.001~\mathrm{mag}$ to $0.05~\mathrm{mag}$; each MC run is iterated
$500$ times.
\par
In principle the MC should be run on the actual data sample; however the large dimension of
future datasets (up to $2000$ supernovae in our simulations) makes this approach impractical.
We chose instead to run the MC on a smaller sample of supernovae, to derive a $\mu$
covariance matrix for this smaller dateset, and to calculate the $\mu$ 
covariance matrix for the larger dataset by interpolating the matrix computed
for the smaller one.
\par
More specifically we proceed as follows:
\begin{enumerate}
\item
Generate a set of supernovae, labelled by $a$, at \rs s $z_a = 0.01 \dots 1.7$ in increments of $0.01$; each
supernova has a stretch $S = 1$ and no extinction so that the only source of variation
in the dataset is the one introduced by the filter zero-point uncertainty.
\item
Fit the light curves of these supernovae and obtain the flux at maximum $f_{0k}$
in filter $k$.
\item
Run the MC: at each realization $v$ a different set of magnitude shifts 
$\mathcal{Z}_k$ is generated, converted to flux, and applied to
$f_{0k}$ as in Equation \ref{eq:magtoflux}.
\item
Fit for the distance moduli $\mu^a{}_k$ at each realization $v$; repeat steps $3$ and $4$ for
$500$ realizations.
\item
At the end of the MC, compute a $\mu$ covariance matrix in the usual way: 
for a pair of supernovae denoted by $a$, $b$:
\be
\label{eq:cov}
(V_{\mathrm{ZP}})_{ab} = 
\langle\mu^a\mu^b\rangle - \langle\mu^a\rangle\langle\mu^b\rangle,
\ee
where  
\begin{eqnarray}
\label{eq:mean}
\langle\mu^a\mu^b\rangle &=& \frac{1}{N_\mathrm{NR}}\sum_{v=1}^{N_\mathrm{NR}}\mu^a{}_v\mu^b{}_v \\
\nonumber
\langle\mu^a\rangle &=& \frac{1}{N_\mathrm{NR}}\sum_{v=1}^{N_\mathrm{NR}}\mu^a{}_v
\end{eqnarray}
and $N_\mathrm{NR} = 500$.
\end{enumerate}
The covariance matrices thus computed are stored for use with the full dataset.
For each pair of supernovae $i$, $j$, at redshifts $z_i$, $z_j$, in the full dataset, an 
entry of $(V_{\mathrm{ZP}})_{ij}$ is computed by spline interpolating the 
matrix defined in Equation \ref{eq:cov} between \rs s $z_a,z_b$ and
$z_{a+1},z_{b+1}$ with $z_a\leqslant z_i\leqslant z_{a+1}$ and
$z_b\leqslant z_j\leqslant z_{b+1}$; this matrix is added to the statistical
$\mu$ covariance matrix, $V_\mu$ and to any other covariance
matrix describing some systematic, $V_\mathrm{Sys}$, such as the
systematic model described in \citet{linder03}.
\par
\begin{figure}
\begin{center}
\epsscale{1.5}
\plottwo{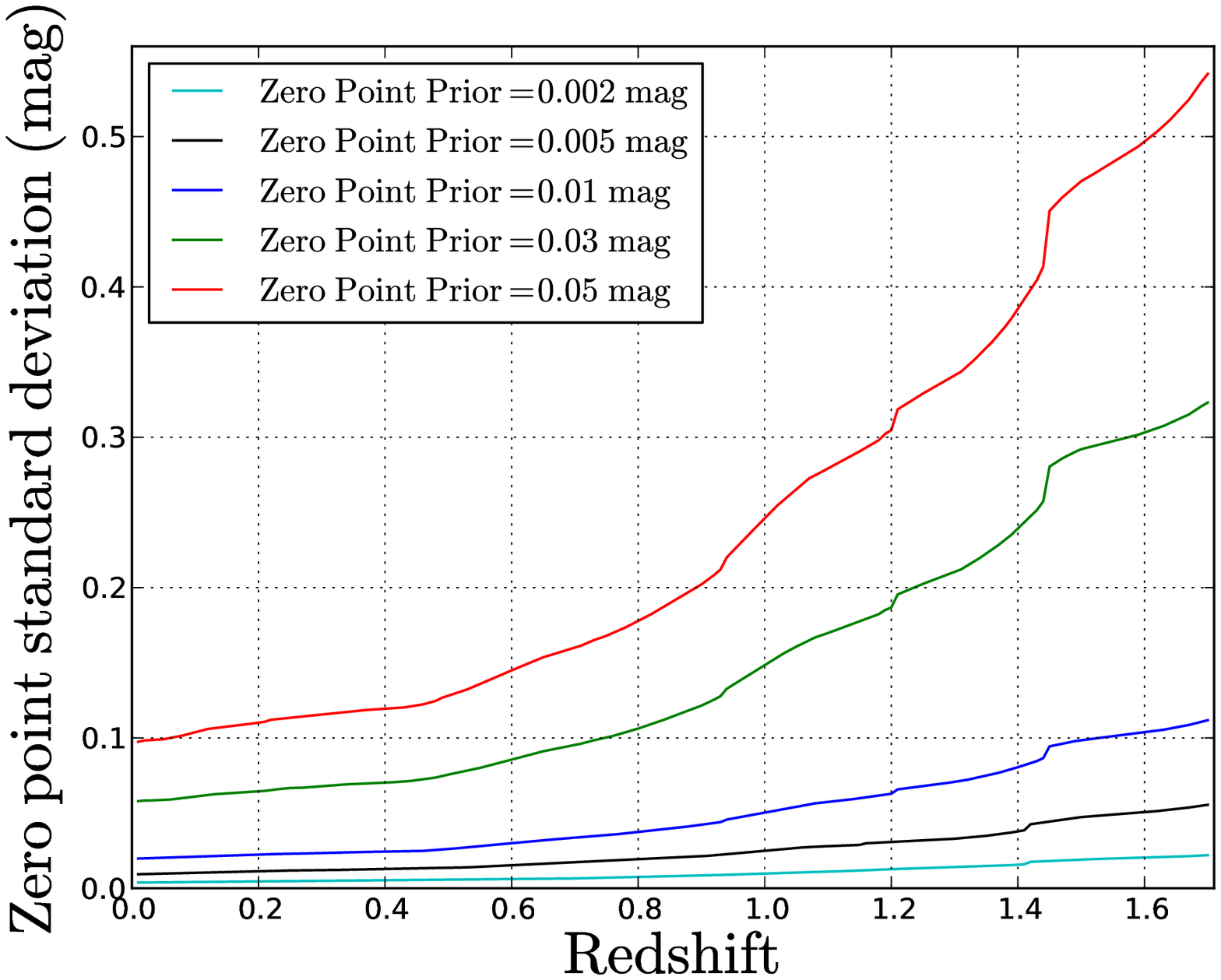}{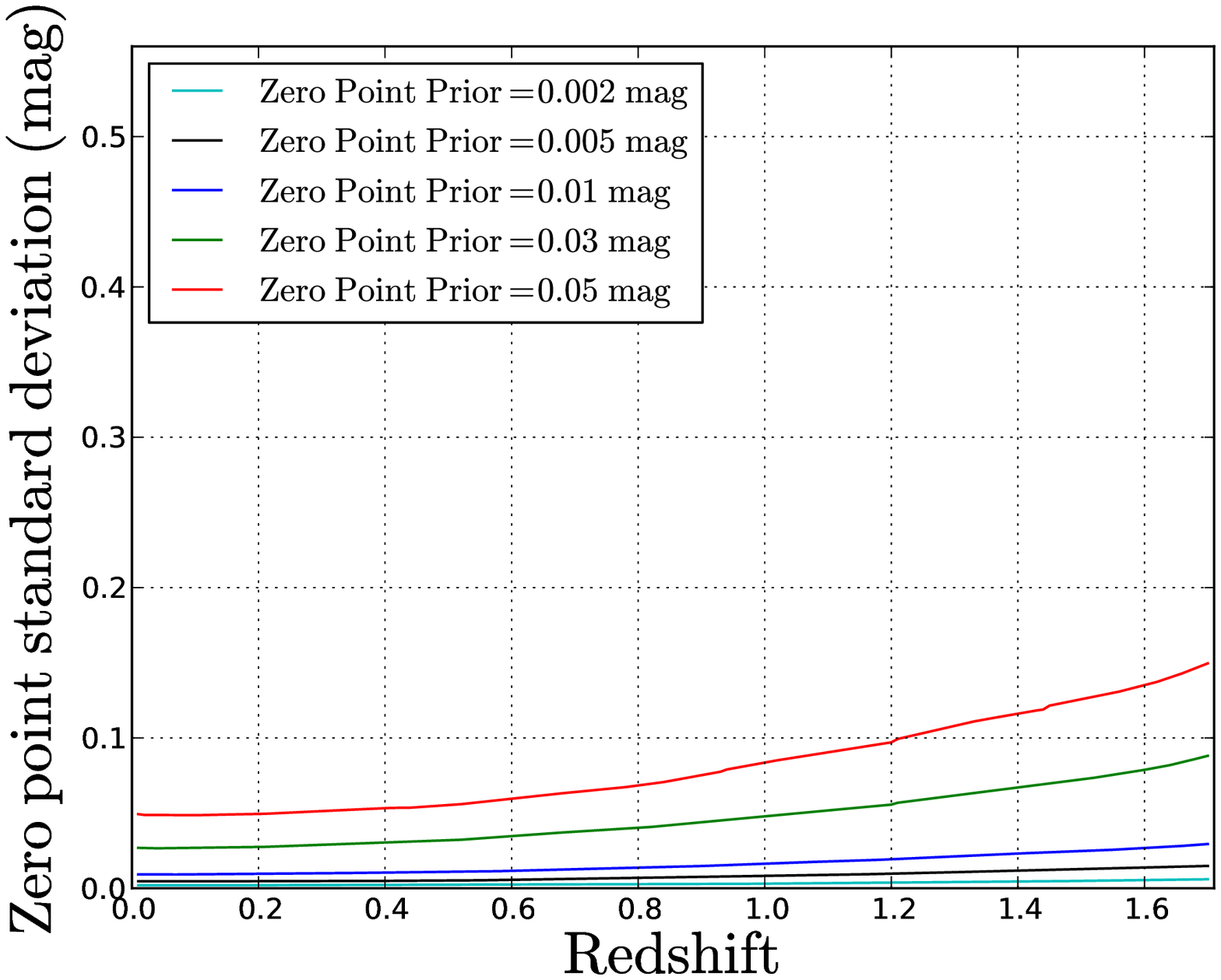}
\end{center}
\caption{Zero-point covariance matrices standard deviations $\sigma_a$ derived from MC, for different values of the zero-point priors, as a
function of redshift.
Upper panel: fitting for $B_V\equiv A_V / R_V$.
Lower panel: fixing $R_V = 3.1$.
The two panels have the same scale to show how fitting for $B_V$ significantly increases $\sigma_a$.}
\label{fig:diagcovz}
\end{figure}

Figure \ref{fig:diagcovz} shows, in the upper panel, the square root of the diagonal elements of the covariance matrix
computed via Equations \ref{eq:mean}, $\sigma_a\equiv\sqrt{\langle(\mu^a)^2\rangle - \langle\mu^a\rangle^2}$,
for different values of the zero-point prior $\sigma_\mathcal{Z}$.
The high values of these elements compared with the values of the prior, especially at high $z$ (e.g:
$\sigma_a\approx 0.5~\mathrm{mag}$ at $z\approx 1.6$ for $\sigma_\mathcal{Z}=0.05~\mathrm{mag}$)
is explained by the dust model we adopted: a CCM model in which we fit both for $A_V$ and for $B_V\equiv A_V / R_V$.
As an alternative one could fix $R_V$ and fit only for $A_V$ when running the MC; we tried this as well, fixing $R_V = 3.1$,
and obtaining values of $\sigma_a$ a factor of $3.5$ lower than those obtained when fitting for $B_V$; these results are
shown in the lower panel of Figure \ref{fig:diagcovz}, plotted on the same scale as the upper panel to show the difference.
In the results we report later for the SN by SN fit we always use covariance matrices obtained by fitting $B_V$ in the MC.
In both cases the curves are roughly proportional to each other by the ratio of their zero-point priors.
\subsection{Simulating the zero-point uncertainty in the simultaneous fit}
The model we adopt to describe our simultaneous $\mu$ fit, taken from KM, is the following:
for $N$ observed supernovae, with supernova $i$ observed in a set of filters $k=1\cdots N_\mathrm{F}$, we have:
\begin{eqnarray}
\label{eq:simfit2}
m^1_{k} &=& \mu^1 + \alpha(S^1 - 1) + M(z_1)_k + A_V{}^1 a(z_1)_k + B_V{}^1 b(z_1)_k + \mathcal{Z}_{k}\nonumber
\\
s^1{}_{k} &=& S^1\nonumber
\\
&\vdots&\nonumber
\\
m^N_{k} &=& \mu^N + \alpha(S^N - 1) + M(z_N)_k + A_V{}^N a(z_N)_k + B_V{}^N b(z_N)_k + \mathcal{Z}_{k}\nonumber
\\
s^N{}_{k} &=& S^N\nonumber
\\
\mathcal{Z}^\mathrm{obs}{}_{k} &=& \mathcal{Z}_{k},
\end{eqnarray}
The meaning of the symbols that also appear in Equations \ref{eq:snmodel} is the same and
the effect of the filter zero-point uncertainty is modelled by the set of parameters
$\mathcal{Z}_k$, one for each filter.
This contrasts with the SN by SN case where the $\mathcal{Z}_k$ are treated as random magnitude shifts chosen from a 
defined probability distribution and added to the peak magnitudes.
$\mathcal{Z}^\mathrm{obs}{}_{k}$ are measured zero-point values and their uncertainty is described by a measurement
covariance matrix $V_{\mathcal{Z}}$; in the following we assume $\mathcal{Z}^\mathrm{obs}{}_{k}=0~\mathrm{mag}$. 
The covariance matrix $V_{\mathcal{Z}}$ may or may not be diagonal; the diagonal case 
$V_{\mathcal{Z}}=\mathrm{diag}(\sigma_1{}^2\cdots\sigma_{N_\mathrm{F}}{}^2)$ (for $N_\mathrm{F}$ filters) amounts of course
to assuming that the filter zero-points uncertainty are all uncorrelated.
This simple assumption is made in KM and, while it is too simplistic, deriving a more realistic model would require a detailed knowledge
of the actual experiment.
In our analysis we will not try to do that but will rather consider uncorrelated zero-points (but see \citet{samsing09} for an
attempt to model the effect of correlated zero-point uncertainties between filters via Principal Component Analysis).
In writing down Equations \ref{eq:simfit2} we have implicitly assumed that the zero-points do not vary in
time and therefore can be represented by a single set of $\mathcal{Z}_{k}$ parameters.
This is a reasonable assumption for the space-based mission we will consider in our simulations, but may not be not be for other
experiments; for example in a ground-based experiment zero-points may be expected to vary with atmospheric conditions.
However, even in the case of time-varying zero-points the KM can still be used; in the ground-based case mentioned above
one may introduce separate sets of $\mathcal{Z}_{k}$ for a set of different atmospheric conditions and assign a set to each
supernova depending on these conditions on the date of observation; other cases where this approach can be used
are modelling changes in the instrument during a very long mission or combining different experiments.
\par
Equations \ref{eq:simfit2} form a linear system of $2\langle N_\mathrm{Obs}\rangle\times N$ observations
and $N_\mathrm{Par}\equiv 4\times N + N_\mathrm{F}$ parameters, where $\langle N_\mathrm{Obs}\rangle$ is the mean number of
observed bands used in the fit per supernova ($\approx 5$ in our simulations), the factor of $2$ is there because for each
band we have a peak magnitude and a stretch, and the factor of $4$ is there because each supernova
is described by four parameters $\mu,S,A_V,B_V$ in addition to the zero-point parameters $\mathcal{Z}_k$.
For a typical stage IV space-based dark energy mission this may translate to $\approx 22000$ observations and $\approx 9000$ parameters;
fortunately the Fisher matrix of the system of Equations \ref{eq:simfit2}, whose inversion is the main computational hurdle in implementing
the KM model, is very sparse since the only non-zero entries come from $4\times 4$ matrices along its diagonal,
corresponding to the supernova parameters, and from the entries whose row or column index correspond to the zero-point
parameters that introduce correlations among the supernovae; therefore the total number of non-zero entries scales as $N$, not $N^2$;
the solution of the system of Equations \ref{eq:simfit2} can therefore be accomplished in about one hour on a 3 GHz desktop with 16 GB
of memory.
\par
Note that when $V_{\mathcal{Z}}\rightarrow 0$ the supernovae in the system of Equations \ref{eq:simfit2}
become decoupled and the model reduces to the traditional SN by SN fit; the cosmology fit results of the two approaches
must then be the same; mathematically the entries in the Fisher matrix whose row or column indices correspond to the zero-point
parameters $\mathcal{Z}_k$ become zero and the Fisher matrix itself becomes block diagonal, with a $4\times 4$ non
zero block for each supernova.
\subsection{Modelling the input data uncertainties}
We have so far focused on the zero-point uncertainties, but other sources of
uncertainty affect the measured magnitudes in each band.
The most important of these are: measurement errors due to Poisson noise
(which was approximated as a Gaussian in our simulation tool),
a possible color uncertainty, any remaining statistical uncertainty, and any
remaining systematic not described by our model.
\par
We model the remaining statistical uncertainties by assuming an intrinsic dispersion
$\sigma_\mathrm{Disp}=0.1~\mathrm{mag}$ for each supernova in the dataset after stretch and color correction;
this value is consistent with values of intrinsic dispersion
used by recent surveys such as ESSENCE \citep{woodvasey07} 
and is also used by the Dark Energy Task Force \citep[DETF:][]{albrecht06} for stage IV experiments such as JDEM.
\par
We then include the possibility of an intrinsic color dispersion, which is not modelled by adding
the same intrinsic dispersion to each supernova, since this affects each band in the same way.
Instead we model an intrinsic color dispersion by adding a new, in principle
non-diagonal covariance matrix to the diagonal Poisson measurement covariance matrix
before fitting for the model described by Equations \ref{eq:simfit2}.
Therefore we have the following model of input data uncertainties:
$V_\mathrm{SN~Data} = V_\mathrm{Poisson} + V_{\delta_c}$.
For simplicity we will consider only $V_{\delta_c}=\mathrm{diag}(\delta_c^2)$ where $\delta_c$ is a constant.
Note that in spite of its form $V_{\delta_c}$ affects supernovae
at different $z$, and therefore observed in a different number of bands, differently:
this color uncertainty model contributes to a magnitude uncertainty of 
$\sim\sqrt{N_\mathrm{obs}}\delta_c$ for a supernova with $N_\mathrm{obs}$ measured bands.
In our analyses we will consider both $\delta_c=0$ and $\delta_c\neq 0$. 
\par
After performing the linear $\mu$ fit described by Equations \ref{eq:simfit2} with $V_{\mathrm{SN~Data}}$
as data covariance matrix, the covariance matrix from the fit, $V_\mu$, is added to the
matrix representing the $0.1~\mathrm{mag}$ supernova intrinsic dispersion described above,
obtaining the matrix $V_\mathrm{Cosmology~Fit}=V_\mu + V_\mathrm{Disp}$,
where $V_\mathrm{Disp}=\mathrm{diag}(0.1~\mathrm{mag})$.
This is then used as the data covariance matrix for the cosmology fit described in Subsection \ref{sec:cosmofit} below.
The cosmology fit is also the place where possible additional systematics are taken into account, by
adding an appropriate covariance matrix to $V_\mathrm{Cosmology~Fit}$ (see Section \ref{sec:cosmofit}).
The SN by SN case is handled slightly differently: for each supernova $i$, the distance modulus $\mu^i$ is derived
independently by fitting the model described by Equations \ref{eq:snmodel}, along with a covariance matrix $V^i{}_\mu$
that does not include the effect of zero-point uncertainties.
The uncertainties in $\mu^i$ derived from $V^i{}_\mu$ are combined in a single diagonal matrix and zero-point
uncertainties are included by adding a non-diagonal matrix obtained by interpolation of the matrix $V_\mathrm{ZP}$,
obtained from the MC, as described in Subsection \ref{sec:mc}; $V_\mathrm{Disp}$ and possibly other systematics 
are then included, and the cosmology fit is performed.
\subsection{Cosmology fit}
\label{sec:cosmofit}
We fit to a flat cosmology with Dark Energy Equation of state (EOS) 
parametrized by $w(a) = w_0 + w_a(1 - a)$ where $a = 1/(1+z)$ 
is the scale factor, with a prior on the reduced distance to the last scattering surface 
($\tilde{d}_{\mathrm{LSS}}$) at $z_\mathrm{LSS} = 1089$ with a $0.2\%$ fractional uncertainty where:
\be
\label{eq:comdist}
\tilde{d}_{\mathrm{LSS}} = \sqrt{\Omega_m h^2}\int_0^{z_{\mathrm{LSS}}}\frac{\mathrm{d}z}
{\sqrt{\Omega_m(1+z)^3+(1-\Omega_m)\exp{(3\int_0^z
\frac{1 + w(z^\prime)}{1 + z^\prime}\mathrm{d}z^\prime)}}}.
\ee
This gives an excellent representation of the expected Planck CMB constraints for combining with supernova data
\citep{linder08,deputter09}.
\par
We chose a fiducial flat $\Lambda\mathrm{CDM}$ cosmology with $\Omega_m=0.3$, consistent with
the value found by \citet{kowalski08} when fitting for such a cosmology.
For presenting our results we use the DETF Figure of Merit \citep[FoM:][]{albrecht06}
as the reciprocal of the square root of the determinant of the covariance matrix after marginalization to
$w_0, w_a$.
This now allows us to investigate the effect of zero-point uncertainties in supernova experiments to understand
dark energy.
\subsection{Mission simulation}
\begin{figure}
\begin{center}
\epsscale{1.1}
\plotone{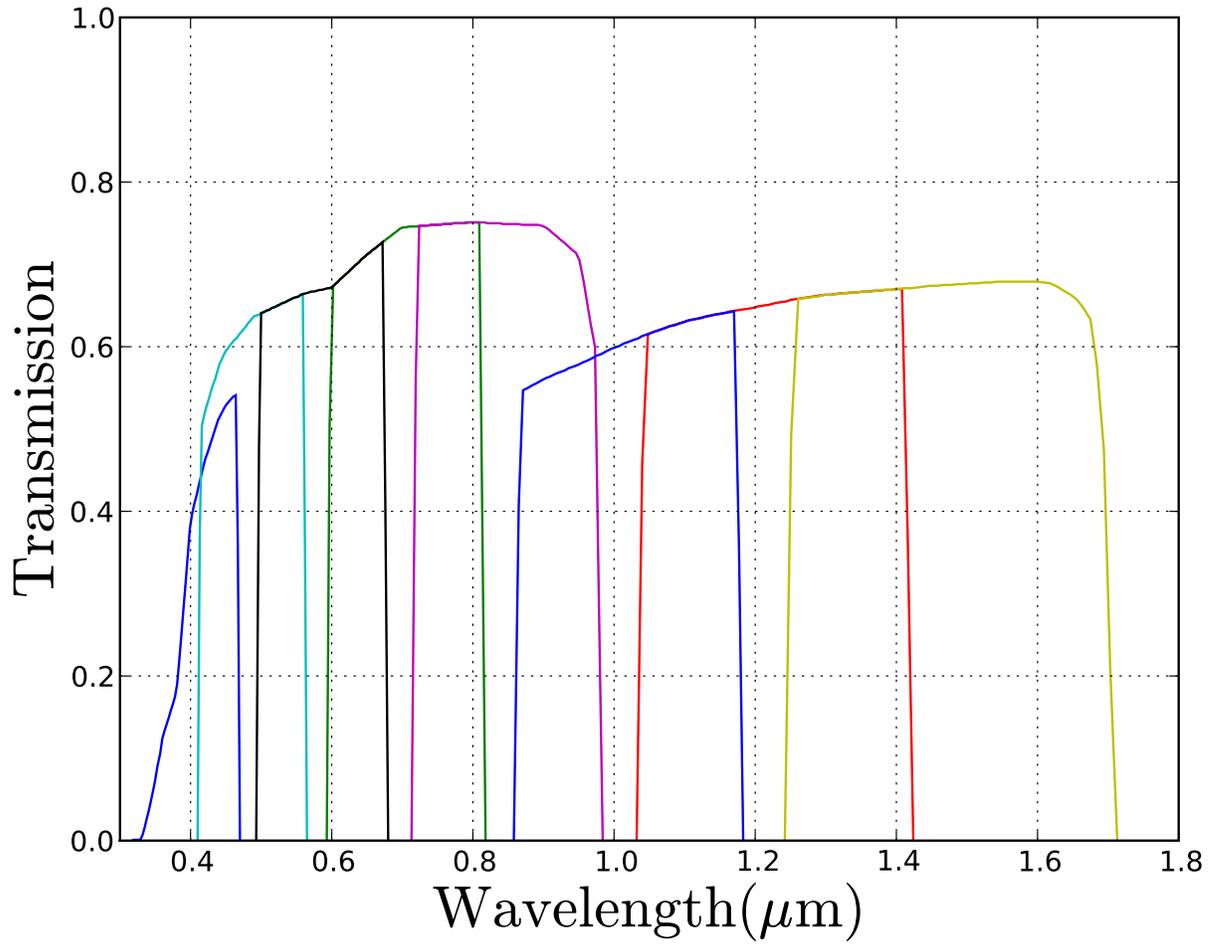}
\end{center}
\caption{Channel throughputs for the eight channels (Five optical and three NIR) 
assumed in the simulation; the transmissions refer to the telescope$+$filter$+$detector combinations.}
\label{fig:channels}
\end{figure}

While we have so far been quite general in describing our zero-point uncertainty model,
we now focus on a specific example of a mission.
We choose to simulate a space mission based on the proposed SuperNova Acceleration Probe\footnote{\url{http://snap.lbl.gov}}
(SNAP) satellite \citep{aldering04}, but with a different configuration than the one described there.
The most important differences with the original SNAP proposal concern the telscope aperture, the number of filters, the maximum survey redshift,
and the redshift distribution; these choices are based (at the time this paper is written)
on what the future JDEM mission may look like.
The most important mission parameters we used are reported in Table \ref{tab:mission}.
\tabletypesize{\footnotesize}
\begin{deluxetable}{|cc|}
\tablewidth{0pc}
\tablecaption{Mission parameters.
\label{tab:mission}}
\startdata
\hline
Telescope aperture & $1.5~\mathrm{m}$ \\
Exposure time & $1200~\mathrm{sec}$ in four dithered exposure of $300~\mathrm{sec}$ each. \\
Cadence & $4~\mathrm{days}$ \\
Filters & $5$ in the optical, $3$ in the NIR \\
Observed SNe Ia & 2000 with flat $z$ distribution\\
\enddata
\end{deluxetable}
The throughputs of the eight channels are shown in Figure \ref{fig:channels}; these are the
transmission of the telescope$+$filter$+$detector combinations.
\section{Results}
\label{sec:results}
We use our simulation tool to explore a larger parameter space than KM.
In particular we want to:
\begin{enumerate}
\item
Compare the two fit methods as 
a function of zero-point uncertainties for a baseline mission, modelled on the SNAP satellite,
with realistic $z$ distributions.
We show that the simultaneous fit greatly outperforms the SN by SN fit; therefore
we will concentrate on the simultaneous fit in the subsequent analyses.
\item
Investigate how the FoM, for the simultaneous fit, varies as mission parameters are
changed; in particular we focus on the effects of
\begin{itemize}
\item
maximum survey redshift $z_\mathrm{max}$.
\item
number of supernovae observed by the mission, $N_\mathrm{SN}$.
\end{itemize}
\item
Include additional systematics.
We will focus on the systematics model described in \citet{linder03}
(hereafter referred to as ``LH systematic'').
\end{enumerate}
\subsection{Making the case for simultaneous fit: results for the baseline mission}
We now present the first main result of the paper: fitting for all supernovae at once vastly outperforms the traditional
SN by SN fitting, in the
sense that the FoM decreases much more slowly with increasing zero-point uncertainties in the
simultaneous fit case.
We present results for realistic mission parameters and four different values of the color uncertainty $\delta_c$:
$\delta_c=0$, $0.005$, $0.01$, and $0.02~\mathrm{mag}$; the zero-points $\mathcal{Z}_k$ are assumed uncorrelated;
we always assume $z_\mathrm{max}=1.5$.
Our results are shown in Table \ref{tab:MCvsSimultaneousFlatZdist}.
\tabletypesize{\footnotesize}
\begin{deluxetable}{ccccccccc}
\tablewidth{0pc}
\tablecaption{FoM of SN by SN fit vs. FoM of simultaneous fit,
for color uncertainty $\delta_c=0$, $0.005$, $0.01$, and $0.02~\mathrm{mag}$.
\label{tab:MCvsSimultaneousFlatZdist}}
\tablehead{
\multicolumn{1}{c}{zero-point} &
\multicolumn{2}{c}{$\delta_c=0~\mathrm{mag}$} &
\multicolumn{2}{c}{$\delta_c=0.005~\mathrm{mag}$} &
\multicolumn{2}{c}{$\delta_c=0.01~\mathrm{mag}$} &
\multicolumn{2}{c}{$\delta_c=0.02~\mathrm{mag}$} \\
\cline{2-9}
\multicolumn{1}{c}{uncertainty} &
\multicolumn{1}{c}{SN by SN fit} &
\multicolumn{1}{c}{Sim. fit} &
\multicolumn{1}{c}{SN by SN fit} &
\multicolumn{1}{c}{Sim. fit} &
\multicolumn{1}{c}{SN by SN fit} &
\multicolumn{1}{c}{Sim. fit} &
\multicolumn{1}{c}{SN by SN fit} &
\multicolumn{1}{c}{Sim. fit} \\
\multicolumn{1}{c}{($\mathrm{mag}$)} &
}
\startdata
0 & 311 & 311 & 306 & 306 & 295 & 295 & 262 & 262\\
0.001 & 246 & 309 & 244 & 302 & 238 & 288 & 217 & 252\\
0.002 & 167 & 309 & 166 & 298 & 163 & 283 & 153 & 246\\
0.003 & 122 & 308 & 121 & 295 & 120 & 278 & 113 & 239\\
0.004 & 95 & 308 & 94 & 292 & 93 & 273 & 88 & 232\\
0.005 & 76 & 308 & 76 & 291 & 75 & 268 & 71 & 226\\
0.006 & 63 & 308 & 62 & 290 & 61 & 265 & 58 & 219\\
0.01 & 35 & 308 & 35 & 288 & 35 & 257 & 33 & 202\\
0.02 & 17 & 308 & 17 & 287 & 17 & 252 & 16 & 186\\
0.03 & 11 & 308 & 11 & 286 & 10 & 251 & 10 & 181\\
0.04 & 7 & 308 & 7 & 286 & 7 & 251 & 7 & 179\\
0.05 & NC\tablenotemark{\dagger} & 308 & NC\tablenotemark{\dagger} & 286 & NC\tablenotemark{\dagger} & 250 & NC\tablenotemark{\dagger} & 179\\
\enddata
\tablenotetext{\dagger}{Fit did not converge.}
\end{deluxetable}
Several things are worth noting in Table \ref{tab:MCvsSimultaneousFlatZdist}:
\begin{enumerate}
\item
In the case of no zero-point uncertainty the two methods give the same result as they must, since
in this case they are mathematically equivalent.
\item
The simultaneous fit vastly outperforms the traditional SN by SN both for $\delta_c=0$ and for
the more realistic case $\delta_c\ne 0$.
This point was already made by KM, but it is reassuring to see that we can
confirm their result for a more realistic mission architecture and with a more sophisticated analysis. 
Therefore the numbers in Table \ref{tab:MCvsSimultaneousFlatZdist}
strongly argue for adopting the simultaneous fit
as a general analysis method for future supernova surveys.
\item
For the not very realistic case of $\delta_c=0~\mathrm{mag}$ the FoM for the simultaneous fit is almost flat as the 
zero-point varies.
This is because in this case self-calibration works so well that the $0.1~\mathrm{mag}$ intrinsic dispersion
dominates the error budget.
In the more realistic case of  $\delta_c\ne0$ the simultaneous fit is still superior: the FoM does decline modestly because of the interaction of
$\delta_c$ with $\mathcal{Z}_k$ each of which affects bands rather than supernovae as a whole.
\end{enumerate}
\par
\begin{figure}
\begin{center}
\epsscale{1.1}
\plotone{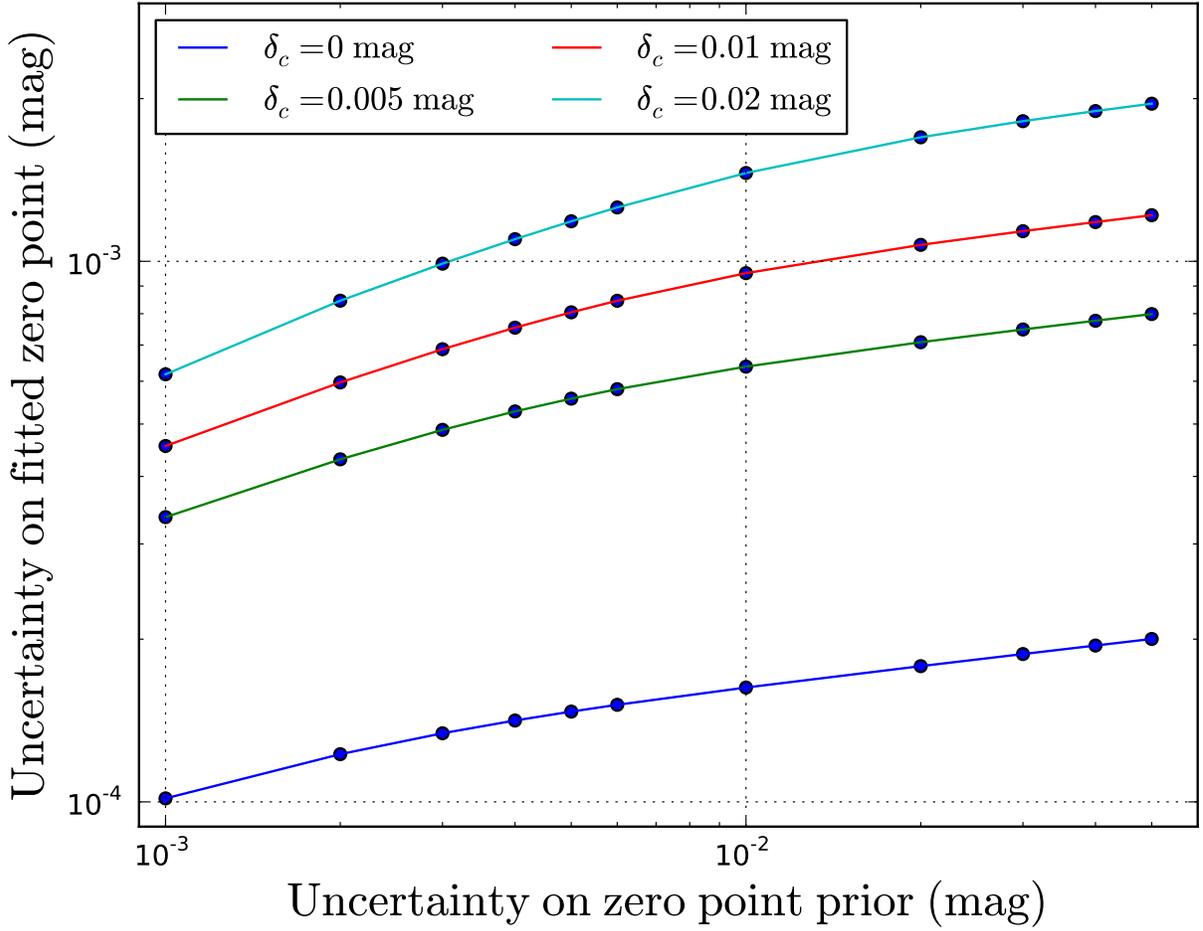}
\end{center}
\caption{Typical statistical uncertainty on the zero point parameters $\mathcal{Z}$ after the fit, $\sigma_{\mathcal{Z}~\mathrm{Fit}}$,
vs. uncertainty on the zero-point prior, $\sigma_\mathcal{Z}$, for color uncertainty $\delta_c=0, 0.005, 0.01, 0.02~\mathrm{mag}$.}
\label{fig:detz}
\end{figure}

To gain more insight into the working of this self-calibration mechanism we consider how the final statistical uncertainties 
on the fit parameters $\mathcal{Z}_k$ are related to the uncertainties on the zero-point priors $\sigma_\mathcal{Z}$.
Quantitatively we consider the subcovariance matrix of the $\mathcal{Z}$ parameters alone
obtained from the $\mu$ fit: its determinant $\mathrm{det}(\mathcal{Z})$ is simply the product of the eigenvalues of this submatrix and
$\mathrm{det}(\mathcal{Z})^{1/N_\mathrm{F}}$, where $N_\mathrm{F}=8$,
should give an estimate of the typical statistical uncertainty in the fit parameters $\mathcal{Z}$ \emph{after} the simultaneous
$\mu$ fit; we call this determinant $\sigma_{\mathcal{Z}~\mathrm{Fit}}$ to emphasize this point.
We compare $\sigma_{\mathcal{Z}~\mathrm{Fit}}$ with the uncertainty on the zero-point prior \emph{before} the fit,
$\sigma_\mathcal{Z}$, in Figure \ref{fig:detz}; the four lines show results for $\delta_c=0, 0.005, 0.01, 0.02~\mathrm{mag}$.
The figure shows that for $\delta_c=0$, $\sigma_{\mathcal{Z}~\mathrm{Fit}}$ grows very slowly as a function of $\sigma_\mathcal{Z}$:
for $\sigma_\mathcal{Z}=0.05~\mathrm{mag}$, $\sigma_{\mathcal{Z}~\mathrm{Fit}}=2\times 10^{-4}~\mathrm{mag}$; this explains the
almost constant FoM as a function of $\sigma_\mathcal{Z}$ for $\delta_c=0~\mathrm{mag}$ reported in Table
\ref{tab:MCvsSimultaneousFlatZdist}.
For $\delta_c\ne 0$ $\sigma_{\mathcal{Z}~\mathrm{Fit}}$ is higher by a factor of $3-5$ than the  $\delta_c = 0~\mathrm{mag}$ case
even at low $\sigma_\mathcal{Z}$ and grows more rapidly as $\sigma_\mathcal{Z}$ increases, but it is still
much smaller than $\sigma_\mathcal{Z}$: for example, for $\delta_c = 0.02~\mathrm{mag}$, at $\sigma_\mathcal{Z}=0.05~\mathrm{mag}$, 
$\sigma_{\mathcal{Z}~\mathrm{Fit}}=2\times 10^{-3}~\mathrm{mag}$, a factor of $10$ higher
than the value for  $\delta_c = 0~\mathrm{mag}$, but more than $10$ times smaller than $\sigma_\mathcal{Z}$.
Therefore Figure \ref{fig:detz} shows both why $\delta_c\ne 0$ decreases the FoM as $\sigma_\mathcal{Z}$ increases
and why the simultaneous fit still outperforms the SN by SN fit.
\par
\begin{figure}
\begin{center}
\epsscale{1.1}
\plotone{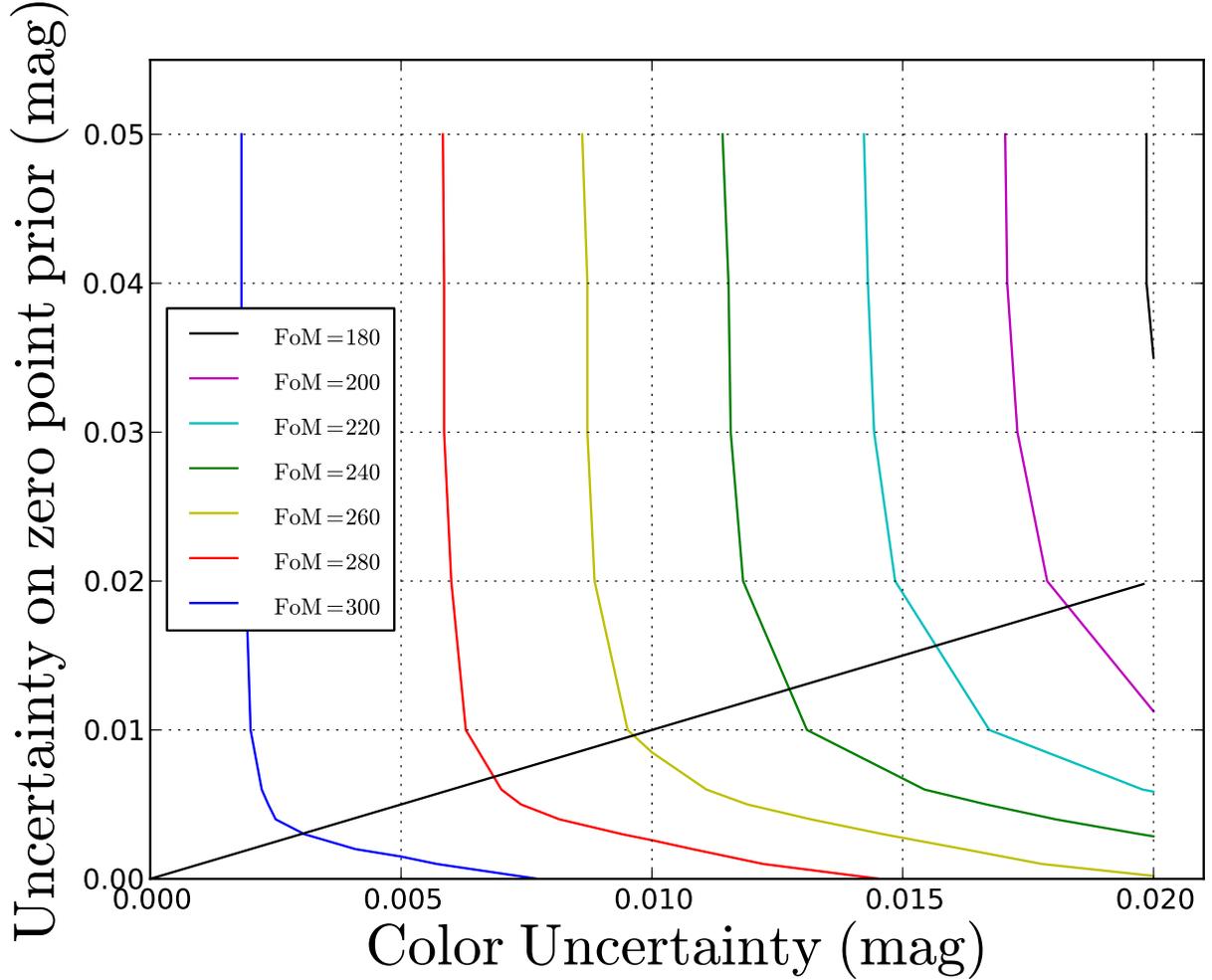}
\end{center}
\caption{Contours of constant FoM for different zero-point prior $\sigma_\mathcal{Z}$ and color uncertainties $\delta_c$.
The straight line at $45^\circ$ shows the points where zero-point prior and color uncertainty are equal; note that the line
intersects the contours roughly where they change their slope from almost vertical ($\sigma_\mathcal{Z} > \delta_c$)
to almost horizontal ($\sigma_\mathcal{Z} < \delta_c$).
The case $\sigma_\mathcal{Z} > \delta_c$ is the self-calibration regime: the data themselves determine the zero-point
more precisely than an accurate zero-point calibration.
In the case  $\sigma_\mathcal{Z} < \delta_c$ a tighter control of zero-point uncertainty is necessary to improve the FoM.
}
\label{fig:color_zeropoint_contour}
\end{figure}

Because of the large parameter space we are exploring it is convenient to visualize our results as contour of constant
FoM as a function of two parameters at the same time.
An interesting combination of parameters to consider is given by the zero-point prior $\sigma_{\mathcal{Z}}$ and the color
uncertainty  $\delta_c$: their relative interplay indicates whether more effort should be expended in calibration or in
understanding supernova colors. 
Figure \ref{fig:color_zeropoint_contour} shows contours of constant FoM as a function of the uncertainty on the
zero-point prior $\sigma_{\mathcal{Z}}$ and the color uncertainty $\delta_c$.
The figure shows, not surprisingly, a trade off between these two parameters.
What is more interesting is the nearly vertical shape exhibited by the graphs, indicating that it pays off to tightly control the 
color uncertainty: for example, to achieve a FoM of $240$, limiting $\delta_c$ to $\lessapprox 0.013~\mathrm{mag}$
results in very lax requirements on the zero-point uncertainty (between $0.01~\mathrm{mag} - 0.05~\mathrm{mag}$), whereas
poorer control of the color uncertainty $\delta_c\gtrapprox 0.013~\mathrm{mag}$ imposes strong requirements on the zero-point 
($\lessapprox 0.01~\mathrm{mag}$); similar considerations hold for other FoMs.
Therefore we see the existence of two regimes: the high $\sigma_{\mathcal{Z}}$ regime where tighter control of color uncertainty is
more important, and the low $\sigma_{\mathcal{Z}}$ regime, where tighter control of zero-point prior is more important;
the transition between these regimes occurs when $\delta_c\approx\sigma_\mathcal{Z}$.
For $\delta_c < \mathcal{Z}_k$ (high $\sigma_{\mathcal{Z}}$ regime) the decline in the FoM is roughly $20(\delta_c / 0.01) \%$
independent of $\mathcal{Z}_k$; in this regime the data themselves determine the zero-point more precisely via self-calibration
and tighter control of color uncertainty lead to further improvements whereas tighter zero-point calibration is not essential.
When $\sigma_\mathcal{Z} < \delta_c$ (low $\sigma_{\mathcal{Z}}$ regime) self-calibration is not dominant and tighter zero-point
calibration is necessary to achieve higher FoMs.
\par
This first conclusion for the baseline mission can therefore be drawn from Table \ref{tab:MCvsSimultaneousFlatZdist}
and Figure \ref{fig:color_zeropoint_contour}: in order to have an impact above self-calibration alone, filter zero-point uncertainties
must be similar to or better than the intrinsic color dispersion.
\section{Exploring the mission parameter space}
\label{sec:parspace}
The second aim of this paper is to explore trades in mission design.
In this section we wish to explore variations in several parameters from the baseline mission presented in Section \ref{sec:results},
analyzing the impact on the results.
In particular we focus on two crucial parameters: the maximum survey redshift $z_\mathrm{max}$ and the number of observed
supernovae $N_\mathrm{SN}$, while keeping the remaining parameters unchanged; we are particularly interested in different 
combinations of parameters that give comparable FoMs.
This is a particularly interesting combination of mission parameters to consider because spectroscopically following up
supernovae at high $z$ is very time consuming since the required time scales as $\approx(1+z)^6$;
the parameters in Table \ref{tab:mission} remain unchanged but the mission duration varies as $z_\mathrm{max}$ and
$N_\mathrm{SN}$ change.
\par
We also wish to consider other sources of systematic uncertainty in addition to zero-point.
A much used model of systematic uncertainty in supernova surveys has been presented by \citet{linder03} who introduce a
redshift-dependent systematic that models, e.g., a non-standard luminosity evolution or time-varying host-galaxy dust extinction.
Their model, which we will refer to as the LH systematic, assigns to each supernova
in a bin of central redshift $z_b$ and total width $0.1$ an equal share in quadrature of an
uncertainty $dm=0.02(1.7/z_\mathrm{max})(1+z_b)/2.7$.
Adopting this model, \citet{linder03} show that maximum survey redshifts of $\gtrapprox 1.5$ are necessary
to convincingly see evidence of a variation in $w$.
The same model is used by \citet{kim04} to describe a generic mission systematic, not necessarily due
to time-varying host-galaxy extinction.
We use the LH systematic in this spirit, namely to describe
any other source of systematic not captured by our zero-point uncertainty model, and we repeat the
same set of simulations described in this subsection with the LH systematic added.
The covariance matrix for the cosmology fit $V_\mathrm{Cosmology~Fit}$ is thus given by:
$V_\mathrm{Cosmology~Fit} = V_\mu + V_\mathrm{Disp} + V_\mathrm{LH}$, and the LH systematic is given by:
\begin{equation}
\label{eq:lh}
dm = 0.01\frac{1+z_b}{2.7}
\end{equation}
that is, we divide the LH systematic for $z_\mathrm{max}=1.7$ by two since we include calibration uncertainty separately.
It is important to note that by adding $V_\mathrm{LH}$ we are implicitly assuming that the LH systematic is uncorrelated
with the other systematics; a more detailed treatment should aim at properly taking into account possible correlations; an
example is described by \citet{amanullah10}.  
\par
We now present our results obtained considering the $N_\mathrm{SN},z_\mathrm{max},\sigma_\mathcal{Z}$ combination,
keeping in turn one of these parameters fixed, and varying the other two.
In all cases we will report tables of FoM and contour plots of constant FoM made from these tables;
all results will be given with and without the LH systematic.
We will also assume in the following  $\delta_c=0.01~\mathrm{mag}$.
When we keep $N_\mathrm{SN}$ fixed we choose  $N_\mathrm{SN}=2000$; when we keep
$z_\mathrm{max}$ fixed we choose $z_\mathrm{max}=1.5$.
\subsection{Influence of maximum survey redshift $z_\mathrm{max}$}
\label{sec:zmaz}
\begin{figure}
\begin{center}
\epsscale{1.8}
\plottwo{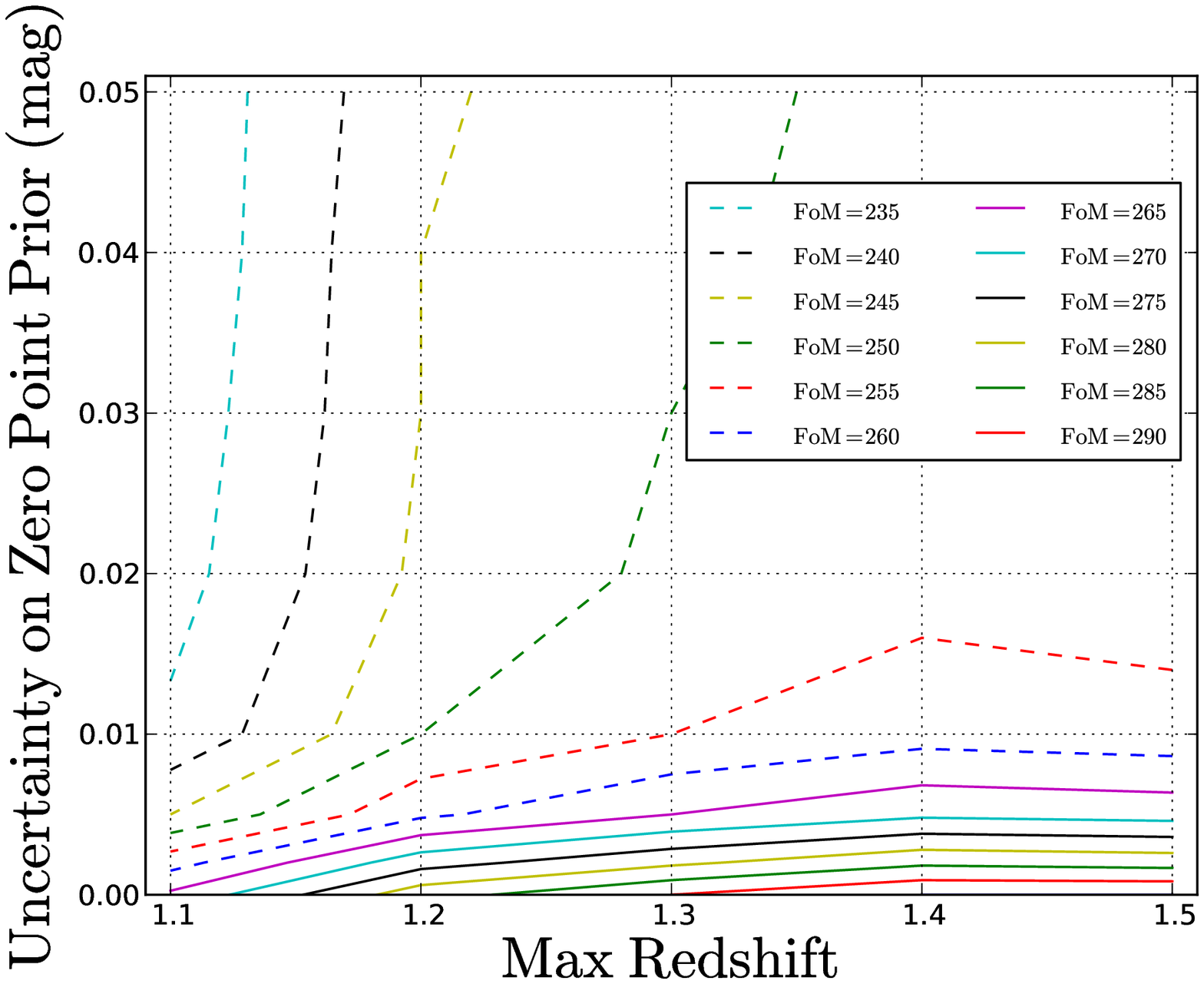}
{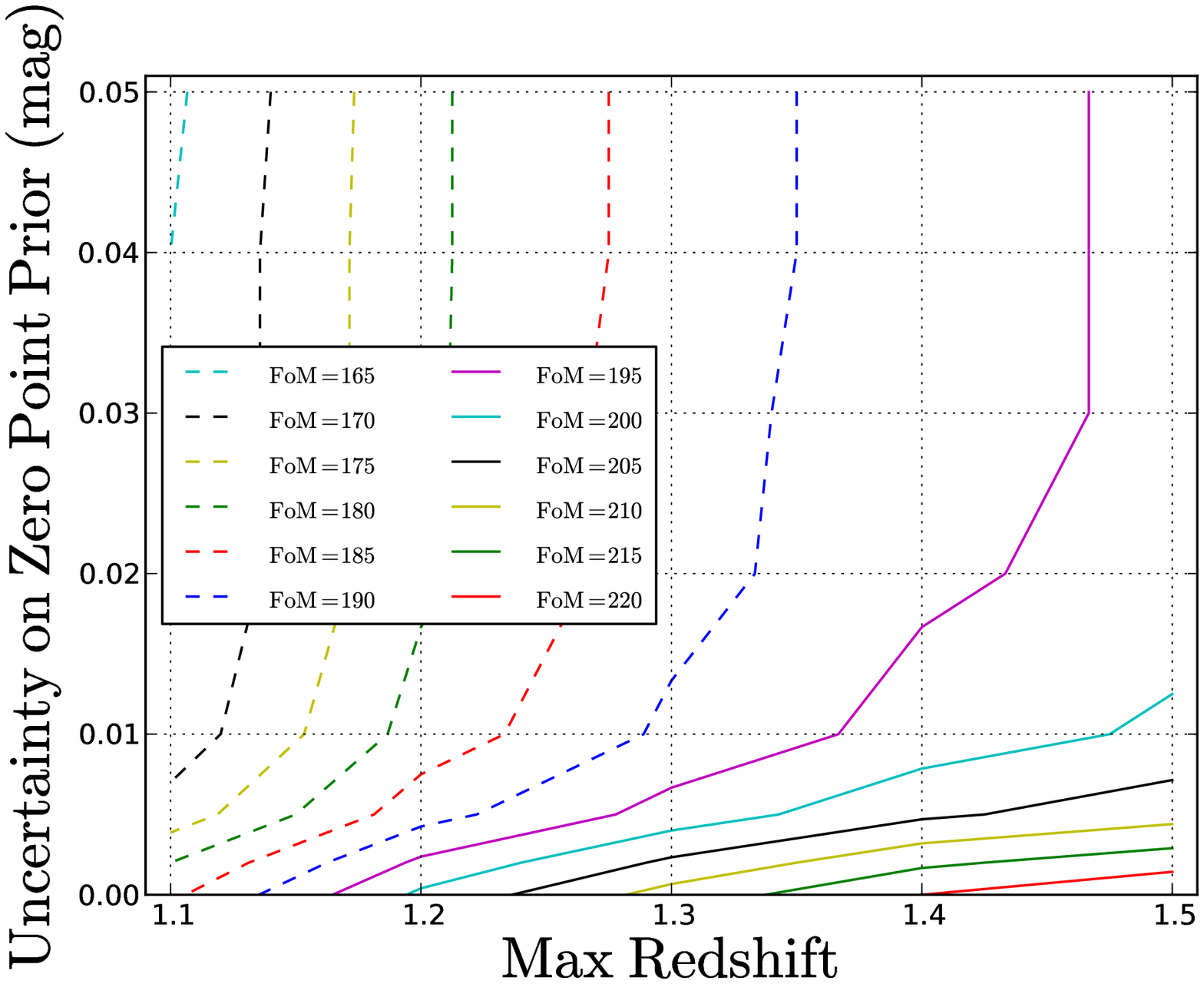}
\end{center}
\caption{Contours of constant FoM for different uncertainties on the zero-point prior $\sigma_\mathcal{Z}$ and maximum survey redshift
$z_\mathrm{max}$.
We assume $N_\mathrm{SN} = 2000$ and $\delta_c = 0.01~\mathrm{mag}$.
Upper panel: the LH systematic is not included.
Lower panel: the LH systematic is included.
}
\label{fig:zmax_zeropoint_contour}
\end{figure}

We consider surveys with  $z_\mathrm{max}=1.1,1.2,1.3,1.4,$ and $1.5$ always keeping the other mission parameters fixed.
The results are reported in Table \ref{tab:zmax2} with and without including the LH systematic;
contour plots of constant FoM as a function of $z_\mathrm{max}$ and $\sigma_\mathcal{Z}$ at constant $N_\mathrm{SN}=2000$
and $\delta_c = 0.01~\mathrm{mag}$ are shown in Figure \ref{fig:zmax_zeropoint_contour} without including the
LH systematic in the upper panel and including it in the lower panel.
\tabletypesize{\footnotesize}
\begin{deluxetable}{cccccc}
\tablewidth{0pc}
\tablecaption{FoM as a function of uncertainty on the zero-point prior for five different $z_\mathrm{max}: 1.1,1.2,1.3,1.4,1.5$.
Upper panel: the LH systematic not included.
Lower panel: the LH systematic included.
In all cases $N_\mathrm{SN}=2000$ and $\delta_c = 0.01~\mathrm{mag}$.
The nearby supernova sample is unchanged.
\label{tab:zmax2}}
\tablehead{
\multicolumn{1}{c}{Uncertainty on zero-point prior $\sigma_\mathcal{Z}$} &
\multicolumn{5}{c}{FoM} \\
\cline{2-6}
\multicolumn{1}{c}{$(\mathrm{mag})$} &
\multicolumn{1}{c}{$z_\mathrm{max} = 1.1$} &
\multicolumn{1}{c}{$z_\mathrm{max} = 1.2$} &
\multicolumn{1}{c}{$z_\mathrm{max} = 1.3$} &
\multicolumn{1}{c}{$z_\mathrm{max} = 1.4$} &
\multicolumn{1}{c}{$z_\mathrm{max} = 1.5$} \\
}
\startdata
0        & 266 & 283 & 290 & 295 & 295 \\
0.002 & 258 & 273 & 279 & 284 & 283 \\
0.005 & 245 & 259 & 265 & 269 & 268 \\
0.010 & 236 & 250 & 255 & 258 & 257 \\
0.02   & 233 & 246 & 251 & 253 & 252 \\
0.03   & 232 & 245 & 250 & 252 & 251 \\
0.04   & 231 & 245 & 249 & 252 & 251 \\
0.05   & 231 & 244 & 249 & 251 & 250 \\
\cline{1-6}
0        & 184 & 201 & 212 & 220 & 225 \\
0.002 & 180 & 196 & 206 & 214 & 218 \\
0.005 & 172 & 188 & 197 & 204 & 208 \\
0.01   & 167 & 182 & 191 & 197 & 201 \\
0.02   & 165 & 179 & 188 & 194 & 197 \\
0.03   & 165 & 179 & 188 & 193 & 196 \\
0.04   & 165 & 179 & 187 & 193 & 196 \\
0.05   & 164 & 179 & 187 & 193 & 196 \\
\enddata
\end{deluxetable}
The upper panel of Figure \ref{fig:zmax_zeropoint_contour} shows the existence of two regimes divided by FoM $\approx 250$:
for FoM $\lessapprox 250$ the contours are almost vertical, wheres for FoM $\gtrapprox 250$ they become almost horizontal.
The former is the self-calibration regime, where, as remarked, the data themselves determine the zero-point precisely;
however self-calibration is less effective with increasing redshift because fewer filter observations are used for each supernova
as $z_\mathrm{max}$ increases.
The figure suggests that if there is no redshift dependent systematic then not surprisingly $z_\mathrm{max}$ becomes less important:
a FoM $=250$ could be achieved for $\sigma_\mathcal{Z} = 0.05~\mathrm{mag}$ and $z_\mathrm{max}=1.3$.
For FoM $>250$ we are not in the self-calibration regime anymore and to achieve FoMs this high
zero-point uncertainties must be tightly controlled ($\sigma_\mathcal{Z}\lessapprox 0.01~\mathrm{mag}$).
\par
The inclusion of the redshift dependent LH systematic however changes the conclusions above as shown in
the lower panel of Figure \ref{fig:zmax_zeropoint_contour}.
Apart from the obvious hit in the FoM it introduces (reducing it by $\approx 70$: clearly the LH systematic is dominant), 
there is also continuous improvement in FoM with higher $z_\mathrm{max}$.
As expected, one can trade off $\sigma_\mathcal{Z}$ and $z_\mathrm{max}$: a FoM $\approx 200$ can be achieved either by
$\sigma_\mathcal{Z}\approx 0.01~\mathrm{mag}$ and $z_\mathrm{max}\approx 1.5$ or
$\sigma_\mathcal{Z}\approx 0.003~\mathrm{mag}$ and  $z_\mathrm{max}\approx 1.3$.
The lower maximum survey redshift, with its reduced spectroscopic time, can achieve similar results \emph{if}
much more stringent zero-point requirements can be met.
\subsection{Influence of the maximum number of observed supernovae $N_\mathrm{SN}$}
\begin{figure}
\begin{center}
\epsscale{1.8}
\plottwo{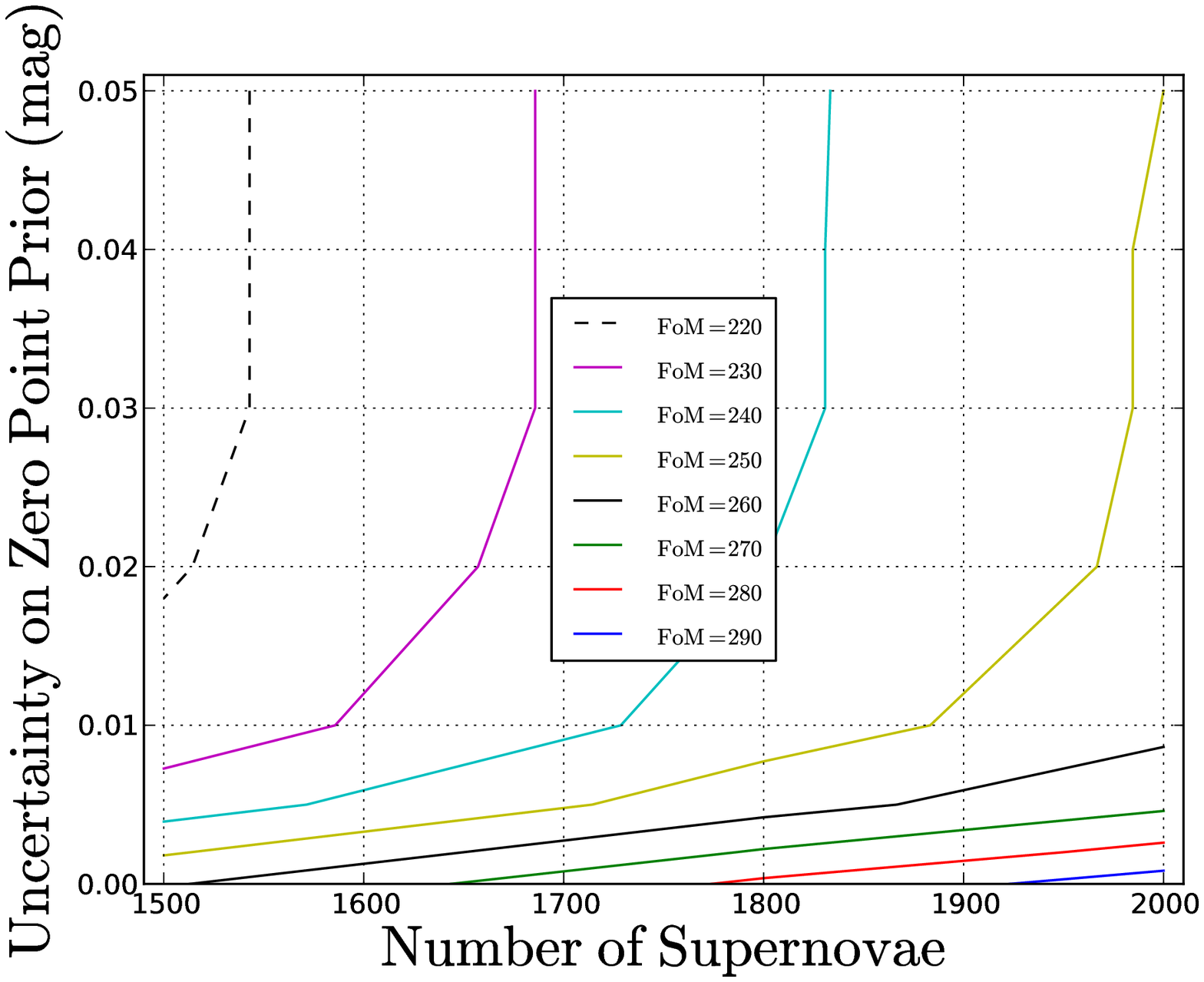}
{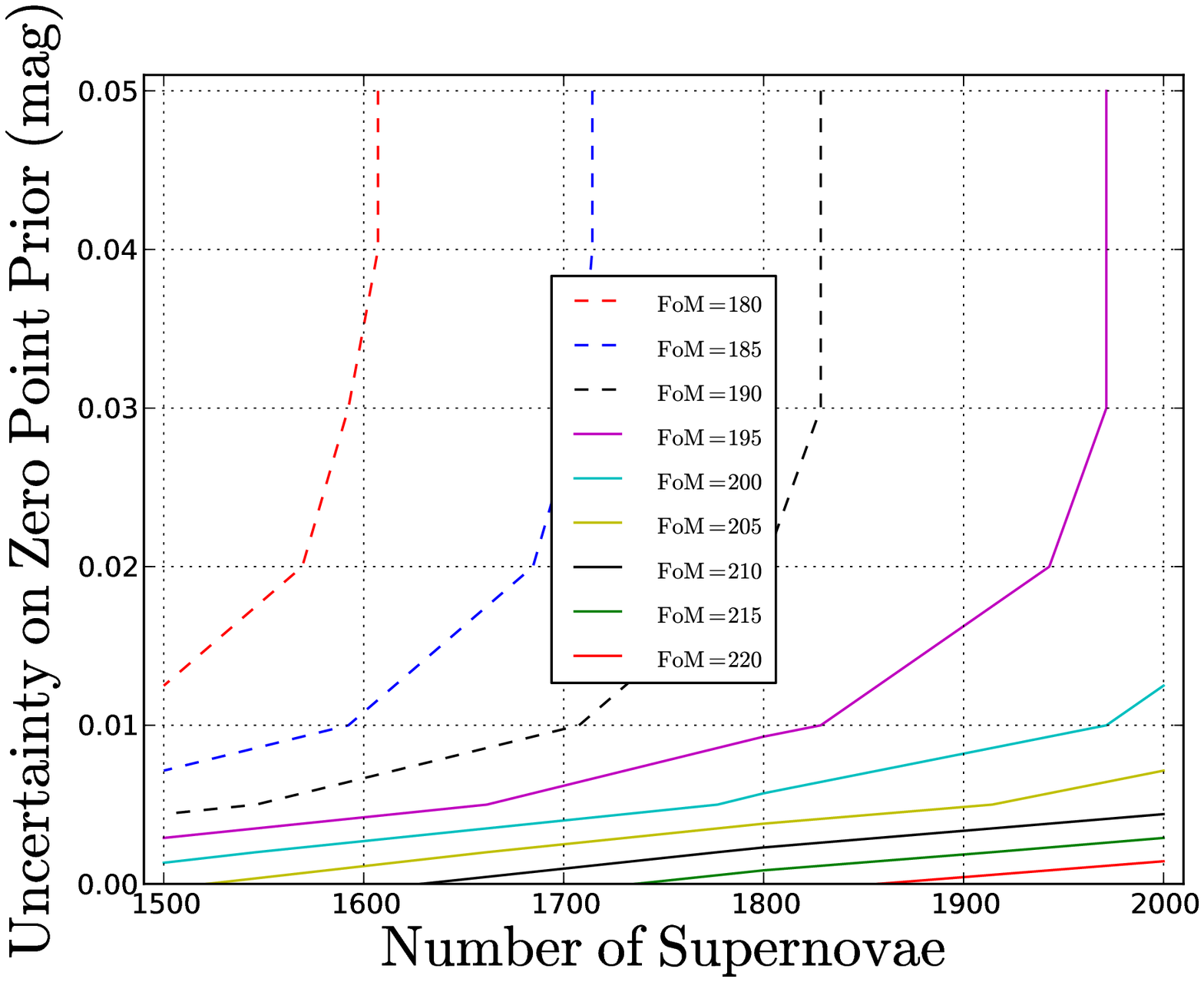}
\end{center}
\caption{Contours of constant FoM for different uncertainties on the zero-point prior $\sigma_\mathcal{Z}$ and
maximum number of supernovae $N_\mathrm{SN}$.
We assume $z_\mathrm{max}=1.5$ and $\delta_c = 0.01~\mathrm{mag}$.
Upper panel: the LH systematic is not included.
Lower panel: the LH systematic is included.
}
\label{fig:nsn_zeropoint_contour}
\end{figure}

We consider surveys with $N_\mathrm{SN} = 1500, 1800, 2000$ while keeping $z_\mathrm{max}=1.5$.
Table \ref{tab:nsn} shows our results with and without including the LH systematic.
Figure \ref{fig:nsn_zeropoint_contour} shows the contour plots made from Table \ref{tab:nsn}
without including the LH systematic in the upper panel and including it in the lower panel.
\tabletypesize{\footnotesize}
\begin{deluxetable}{cccccc}
\tablewidth{0pc}
\tablecaption{FoM as a function of uncertainty on the zero-point prior for three different $N_\mathrm{SN}: 1500,1800,2000$.
Upper panel: the LH systematic is not included.
Lower panel: the LH systematic is included.
In all cases $z_\mathrm{max}=1.5$ and $\delta_c = 0.01~\mathrm{mag}$.
The nearby supernova sample is unchanged.
\label{tab:nsn}}
\tablehead{
\multicolumn{1}{c}{Uncertainty on zero-point prior $\sigma_\mathcal{Z}$} &
\multicolumn{3}{c}{FoM} \\
\cline{2-4}
\multicolumn{1}{c}{$(\mathrm{mag})$} &
\multicolumn{1}{c}{$N_\mathrm{SN} = 1500$} &
\multicolumn{1}{c}{$N_\mathrm{SN} = 1800$} &
\multicolumn{1}{c}{$N_\mathrm{SN} = 2000$} \\
}
\startdata
0       & 259  & 282  & 295 \\
0.002 & 249 & 271  & 283 \\
0.005 & 235 & 256  & 268 \\
0.01  & 224 & 245  & 257 \\
0.02  & 219 & 240  & 252 \\
0.03  & 217 & 238  & 251 \\
0.04  & 217 & 238  & 251 \\ 
0.05  & 217 &238  & 250 \\
\cline{1-4}
0        & 204 & 218 & 225 \\
0.002 & 198 & 211 & 218 \\
0.005 & 188 & 201 & 208 \\
0.01   & 181 & 194 & 201 \\
0.02   & 177 & 190 & 197 \\
0.03   & 176 & 189 & 196 \\
0.04   & 175 & 189 & 196 \\
0.05   & 175 & 189 & 196 \\
\enddata
\end{deluxetable}
In the upper panel of Figure \ref{fig:nsn_zeropoint_contour} we again see the existence of the two regimes distinguished by
FoM $\approx 250$ we noted in Figure \ref{fig:zmax_zeropoint_contour}; this shows that that the larger the number of supernovae per
redshift bin the better the self-calibration can be done.
The figure shows that achieving FoM $\gtrapprox 250$ requires a tight control of
zero-point uncertainties $\sigma_\mathcal{Z}\lessapprox 0.01~\mathrm{mag}$, at least if one considers $N_\mathrm{SN}\le 2000$.
(We did not consider $N_\mathrm{SN} > 2000$ because such numbers would probably be unrealistically high for
a future space-based mission).
For FoM $\lessapprox 250$ on the other hand zero-point requirements are much less severe.
The existence of these two regimes can once again be explained by self-calibration: for FoM~$\lessapprox 250$ and
$\sigma_\mathcal{Z}\gtrapprox 0.02~\mathrm{mag}$ we are in the self-calibration regime and the contours are therefore roughly vertical,
indicating that the FoM is quite insensitive to the actual value of the zero-point prior $\sigma_\mathcal{Z}$.
In this regime it pays to increase $N_\mathrm{SN}$; an increase of $\approx 10$ in FoM can be achieved by observing $\approx 150$ more
supernovae, almost regardless of $\sigma_\mathcal{Z}$.
For FoM $\gtrapprox 250$ and $N_\mathrm{SN}\le 2000$ we are not in the self-calibration regime anymore and the contours are
almost flat: a tighter control of zero-point uncertainties is necessary to achieve higher FoMs.
Figure \ref{fig:nsn_zeropoint_contour} shows that in this regime an increase of $\approx 500$ supernovae, from $1500$ to $2000$
results in only less than $\approx 0.01~\mathrm{mag}$ relaxation in the $\sigma_\mathcal{Z}$ requirement.\par
Including the LH systematic does not change this conclusion much: from the lower panel of
Figure \ref{fig:nsn_zeropoint_contour} we again notice
the overall decrease of about $\approx 70$ in FoM and we see that a tight control of zero-point uncertainties
($\sigma_\mathcal{Z}\lessapprox 0.01~\mathrm{mag}$) is required to achieve FoM $\gtrapprox 200$.
Interestingly, for $\sigma_\mathcal{Z}\gtrapprox 0.02~\mathrm{mag}$, to achieve an increase in FoM $\approx 10$, additional
$\approx 220$ more supernovae are required (at least if uniformly distributed), compared with $\approx 150$ without including
the LH systematic; this conclusion argues for observing modest numbers of supernovae at high $z$ rather than many at lower
$z$, consistent with the conclusions of \citet{linder03}.
\subsection{Varying $N_\mathrm{SN}$ and $z_\mathrm{max}$ simultaneously}
\begin{figure}
\begin{center}
\epsscale{1.8}
\plottwo
{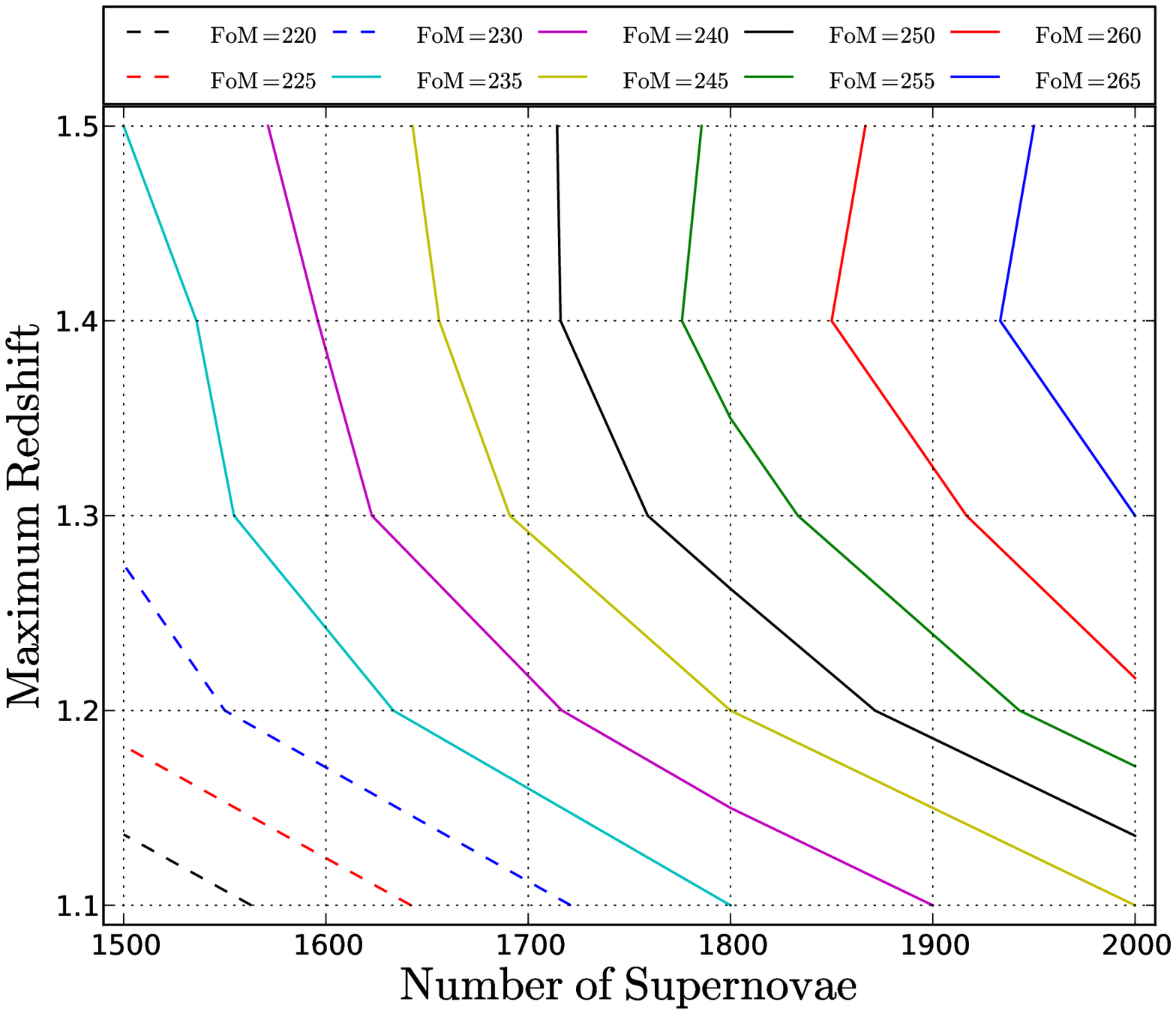}
{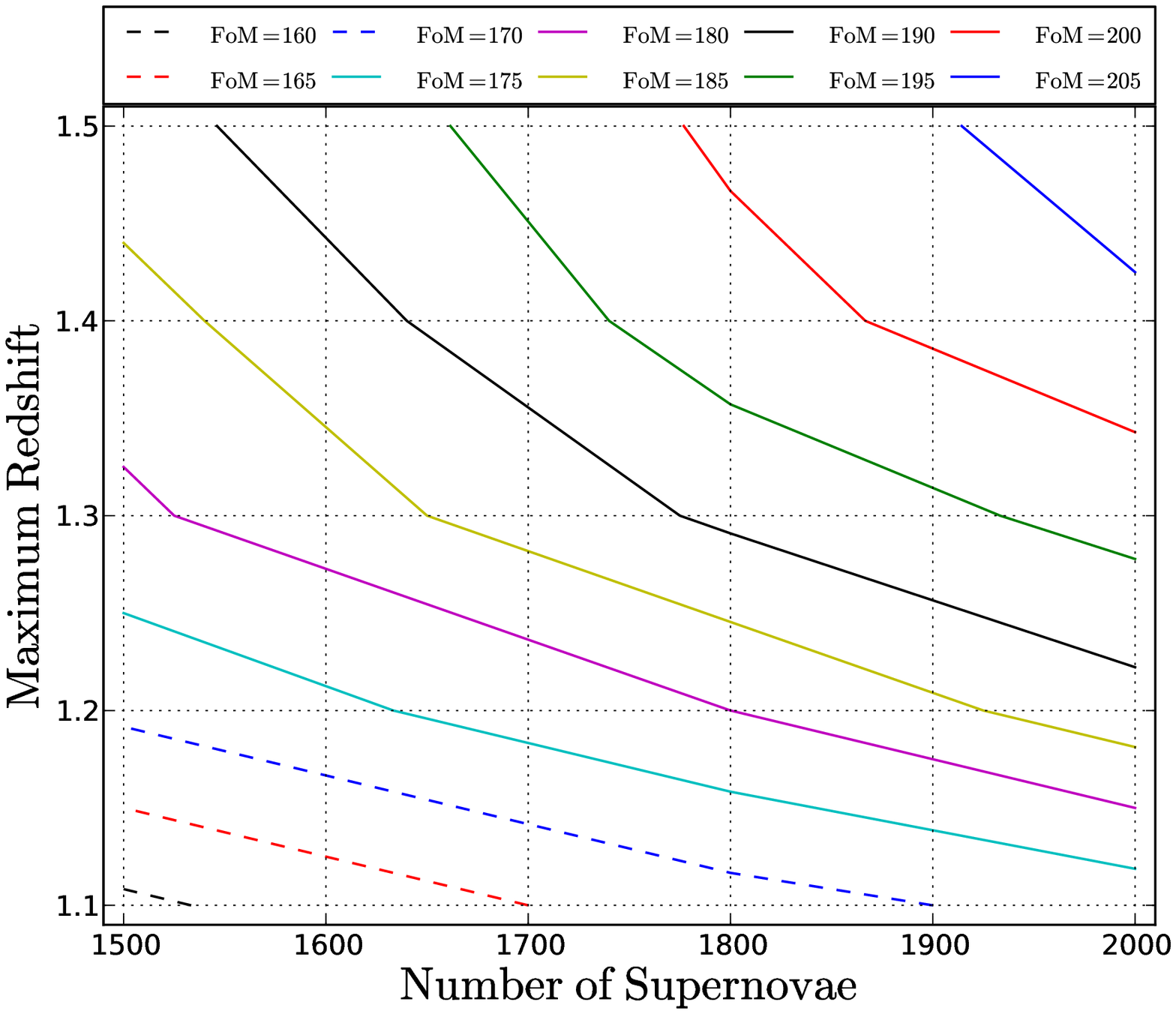}
\end{center}
\caption{Contours of constant FoM for different numbers of supernovae $N_\mathrm{SN}$ and maximum survey redshift $z_\mathrm{max}$
at fixed zero-point prior uncertainty $\sigma_\mathcal{Z}=0.005~\mathrm{mag}$.
We assume $\delta_c=0.01~\mathrm{mag}$.
Upper panel: the LH systematic is not included.
Lower panel: the LH systematic included.}
\label{fig:nsn_zmax_contour_ZP_0.005}
\end{figure}

\begin{figure}
\begin{center}
\epsscale{1.8}
\plottwo
{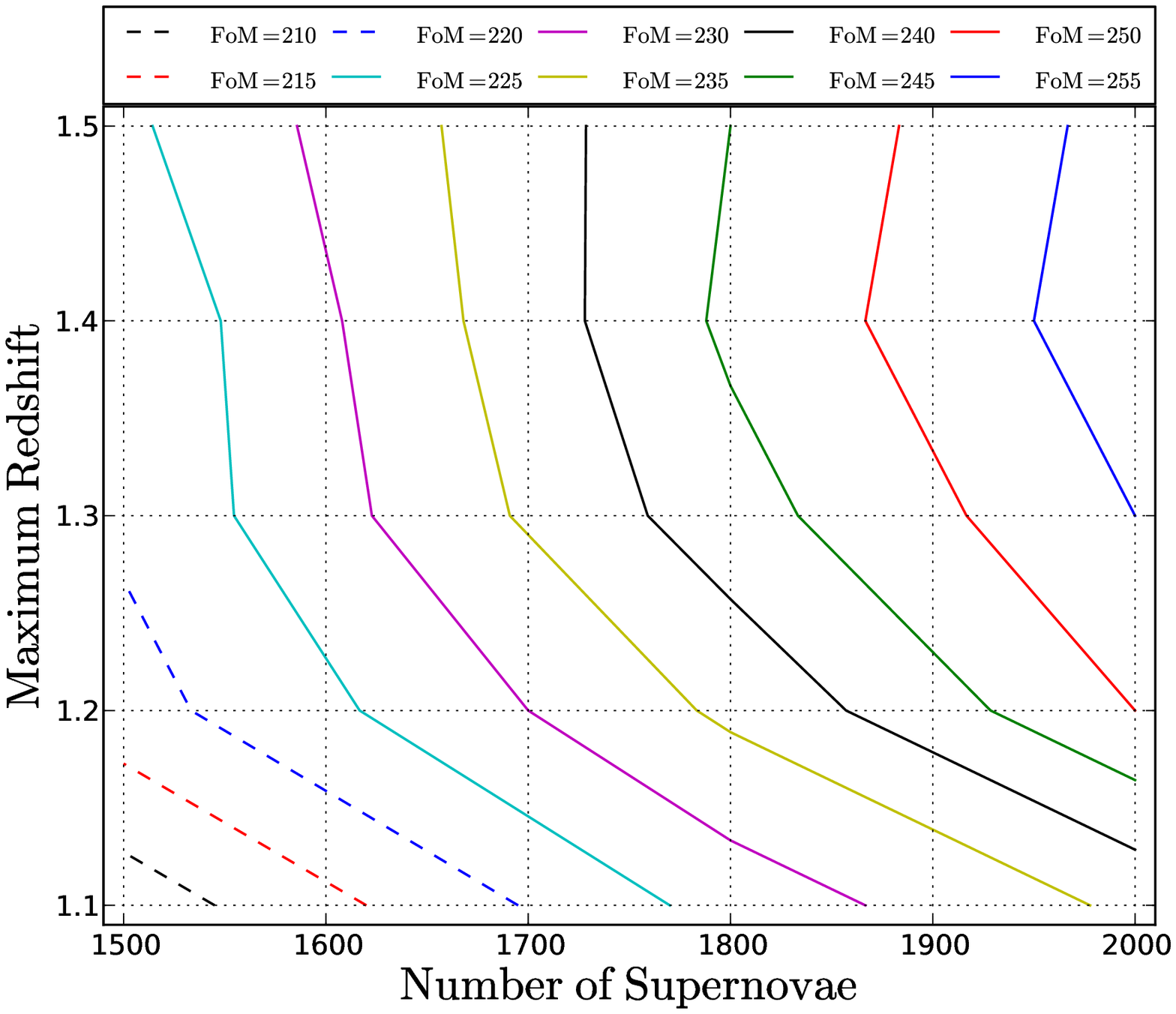}
{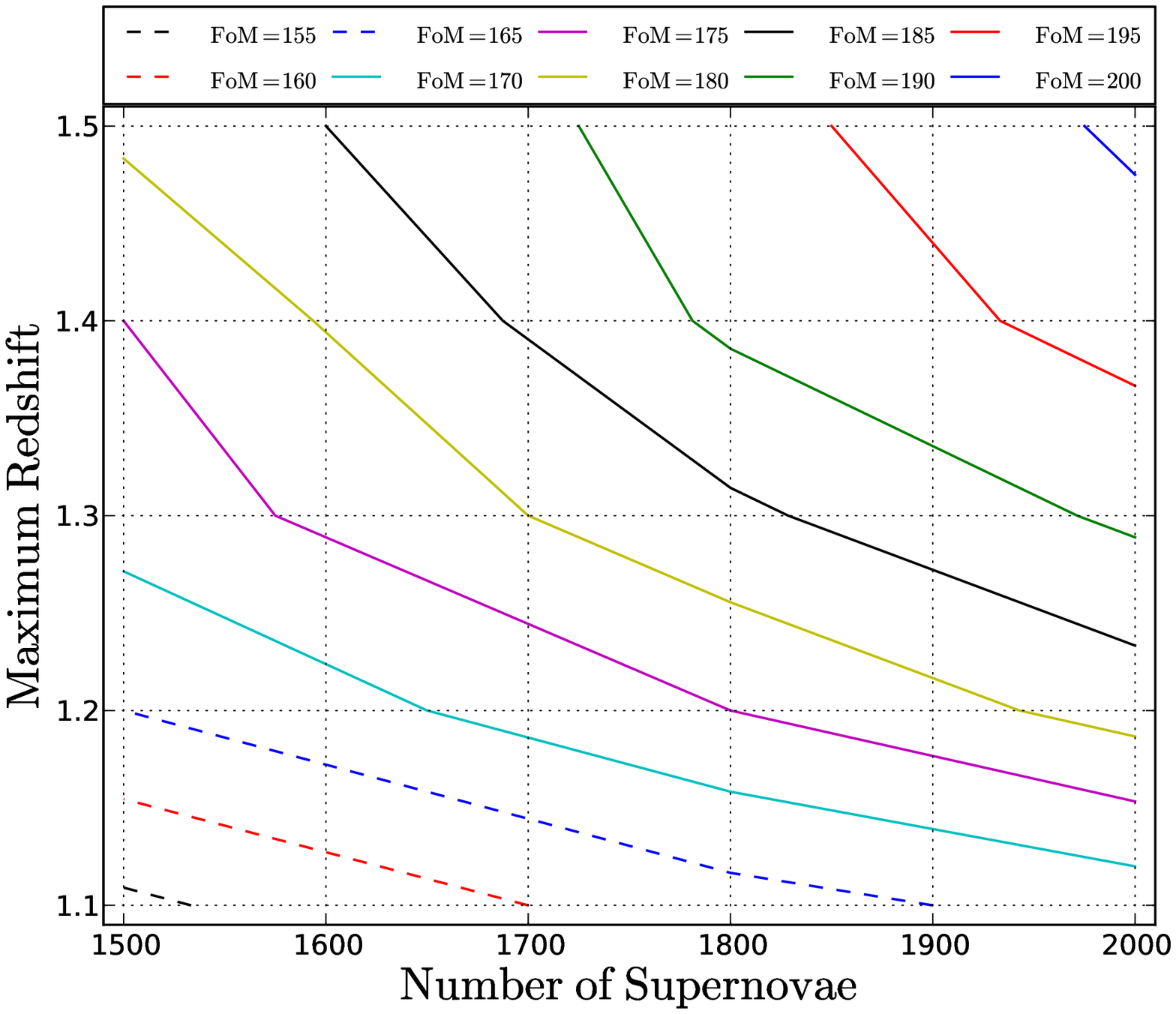}
\end{center}
\caption{Contours of constant FoM for different numbers of supernovae $N_\mathrm{SN}$ and maximum survey redshift $z_\mathrm{max}$
at fixed zero-point prior uncertainty $\sigma_\mathcal{Z}=0.01~\mathrm{mag}$.
We assume $\delta_c=0.01~\mathrm{mag}$.
Upper panel: the LH systematic not included.
Lower panel: the LH systematic included.}
\label{fig:nsn_zmax_contour_ZP_0.01}
\end{figure}

To further investigate the interplay of $N_\mathrm{SN}$ and $z_\mathrm{max}$, we varied them
simultaneously while keeping the uncertainty on the zero-point prior $\sigma_\mathcal{Z}$ fixed at $0.005$ and
$0.01~\mathrm{mag}$.
Again the inclusion of LH systematic changes the conclusion in each case.
\tabletypesize{\footnotesize}
\begin{deluxetable}{cccccc}
\tablewidth{0pc}
\tablewidth{0pc}
\tablecaption{
FoM as a function of number of supernovae $N_\mathrm{SN}$ and maximum survey redshift $z_\mathrm{max}$
for fixed uncertainty on the zero-point prior $\sigma_\mathcal{Z}=0.005,0.01~\mathrm{mag}$.
FoMs are reported both with and without including the LH systematic.
We assume $\delta_c = 0.01~\mathrm{mag}$.
The nearby supernova sample is unchanged.
\label{tab:nsn_vs_zmax}}
\tablehead{
\multicolumn{1}{c}{} &
\multicolumn{1}{c}{} &
\multicolumn{4}{c}{FoM} \\
\cline{3-6}
\multicolumn{1}{c}{$N_\mathrm{SN}$} &
\multicolumn{1}{c}{$z_\mathrm{max}$} &
\multicolumn{1}{c}{$\sigma_\mathcal{Z}=0.005~\mathrm{mag}$} &
\multicolumn{1}{c}{$\sigma_\mathcal{Z}=0.005~\mathrm{mag}$} &
\multicolumn{1}{c}{$\sigma_\mathcal{Z}=0.01~\mathrm{mag}$} &
\multicolumn{1}{c}{$\sigma_\mathcal{Z}=0.01~\mathrm{mag}$} \\
\multicolumn{1}{c}{} &
\multicolumn{1}{c}{} &
\multicolumn{1}{c}{No LH} &
\multicolumn{1}{c}{LH} &
\multicolumn{1}{c}{No LH} &
\multicolumn{1}{c}{LH} \\
}
\startdata
1500 & 1.1 & 216 & 159 & 207 & 154 \\
1500 & 1.2 & 227 & 171 & 218 & 165 \\
1500 & 1.3 & 231 & 179 & 221 & 172 \\
1500 & 1.4 & 232 & 183 & 221 & 175 \\
1500 & 1.5 & 235 & 188 & 224 & 181 \\
\cline{1-6}
1800 & 1.1 & 235 & 168 & 227 & 163 \\
1800 & 1.2 & 245 & 180 & 236 & 175 \\
1800 & 1.3 & 253 & 191 & 243 & 184 \\
1800 & 1.4 & 257 & 198 & 246 & 191 \\
1800 & 1.5 & 256 & 201 & 245 & 193 \\
\cline{1-6}
2000 & 1.1 & 245 & 172 & 236 & 167 \\
2000 & 1.2 & 259 & 188 & 250 & 182 \\
2000 & 1.3 & 265 & 197 & 255 & 191 \\
2000 & 1.4 & 269 & 204 & 258 & 197 \\
2000 & 1.5 & 268 & 208 & 257 & 201 \\
\enddata
\end{deluxetable}
The results are shown in Table \ref{tab:nsn_vs_zmax}; the contour plots drawn from the data in the table are shown in Figures 
\ref{fig:nsn_zmax_contour_ZP_0.005} and \ref{fig:nsn_zmax_contour_ZP_0.01}; the upper panels show results without including the
LH systematic, the lower panels including it.
Both panels in these figures  show, unsurprisingly, a tradeoff between $N_\mathrm{SN}$ and $z_\mathrm{max}$.
Without any redshift dependent systematic, for the higher FoMs (FoM $\gtrapprox 250$ for $\sigma_\mathcal{Z}=0.005~\mathrm{mag}$,
FoM $\gtrapprox 245$ for $\sigma_\mathcal{Z}=0.01~\mathrm{mag}$) going to higher $z_\mathrm{max}$ is not only ineffective, but even
counterproductive: the optimum  $z_\mathrm{max}$ is about $1.4$.
The inclusion of the LH systematic, shown in the lower panels of the figures, changes this conclusion, showing once again the importance of a
high $z_\mathrm{max}$: only high $z_\mathrm{max}$ can achieve high FoM.
Only if one is willing to settle for low FoM can one lower $z_\mathrm{max}$ and compensate by an increase in $N_\mathrm{SN}$.
We finally note that with the simultaneous fit calibration uncertainties are a subdominant component to LH in the error budget, whereas with the
the SN by SN fit calibration uncertainties are dominant.
\section{Discussion and conclusion}
\label{sec:conclusion}
Adopting the general method of modelling zero-point uncertainties introduced by KM we have carried out
simulations of a future space-based supernova dark energy experiment with the main aim of
assessing the influence of zero-point uncertainties on its overall performance.
We have confirmed KM results for a more realistic experiment:
fitting for all supernovae at once results in a greatly improved mission performance over the traditional SN by SN fitting.
Whereas this effect may not be evident in today's surveys involving a few hundreds of supernovae and few available bands,
it will become very significant for future surveys.
We explored a representative section of the mission parameter space paying particular attention to how zero-point requirements can be
traded off with other mission parameters; in particular we have shown that in general going to higher redshift results in less stringent zero-point 
requirements, even without considering other form of systematic.
We stress once again that, while our results are for a specific possible space-based mission, the KM model itself is more general.
The inclusion of a redshift dependent systematic such as the LH systematic greatly affects the mission performance,
both by significantly degrading the FoM and by making the case for higher redshift even stronger; it is therefore extremely
important to better characterize other forms of systematic by the time future stage IV experiments get under way.
Finally the tools used here can realistically simulate future dark energy supernova experiments.
The work can be expanded in many ways, all easily implementable in our simulation tool.
The most obvious examples are different mission architectures, both ground and space-based,
different redshift distributions, further models of systematics.
For the zero-point uncertainties one can explore tighter characterization in the optical vs the near infrared and variation with time.
The latter may be especially relevant for ground-based surveys.
A more detailed treatment of the nearby supernova sample would introduce a separate set of zero-point parameters;
again this can be accommodated by the KM model.
\acknowledgements
This work was supported by the Director, Office of Science, Office of High Energy Physics,
of the U.S. Department of Energy under Contract No. DE-AC02-05CH11231.

\end{document}